\documentclass[twocolumn,aps,pra,superscriptaddress,nofootinbib]{revtex4-1}
\usepackage{graphicx}
\usepackage{dcolumn}
\usepackage{bm}
\usepackage[latin5]{inputenc}
\usepackage{amssymb,amsmath,amsthm,amsbsy,mathptmx}
\DeclareMathAlphabet{\mathcal}{OMS}{cmsy}{m}{n}
\usepackage{tabularx}
\usepackage{xcolor}
\usepackage{mathtools}
\usepackage{oubraces}
\usepackage{dsfont}
\usepackage[title, titletoc]{appendix}
\usepackage{graphicx}
\usepackage{rotating}
\usepackage[most]{tcolorbox}
\usepackage{enumerate}
\usepackage{subfloat}
\usepackage{multirow}
\usepackage{balance}
\usepackage{lastpage}
\usepackage[colorlinks=true,linkcolor=blue,citecolor=blue,urlcolor=blue]{hyperref}%
\usepackage{cleveref}
\Crefname{subfigures}{figure}{figures}
\Crefname{subfigures}{Figure}{Figures}

\newcommand\Tstrut{\rule{0pt}{2.6ex}}         
\newcommand\Bstrut{\rule[-0.9ex]{0pt}{0pt}}   

\newcommand{\nc}{\newcommand}
\nc{\rnc}{\renewcommand}
\nc{\beg}{\begin{equation}}
\nc{\eeq}{{\end{equation}}}
\nc{\beqa}{\begin{eqnarray}}
\nc{\eeqa}{\end{eqnarray}}
\nc{\lbar}[1]{\overline{#1}}
\nc{\bra}[1]{\langle#1|}
\nc{\ket}[1]{|#1\rangle}
\nc{\ketbra}[2]{|#1\rangle\!\langle#2|}
\nc{\braket}[2]{\langle#1|#2\rangle}

\begin{document}

\title{Resource theory of superposition: State transformations}
\author{G\"{o}khan Torun}
\email{gokhantorun62@gmail.com}
\affiliation{Department of Physics, Bo{\u{g}}azi{\c{c}}i University, 34342 Bebek, \.{I}stanbul, Turkey}
\author{H\"{u}seyin Talha \c{S}enya\c{s}a}
\email{senyasa@itu.edu.tr}
\affiliation{Department of Physics, \.{I}stanbul Technical University, 34469 Maslak, \.{I}stanbul, Turkey}
\author{Ali Yildiz}
\email{yildizali2@itu.edu.tr}
\affiliation{Department of Physics, \.{I}stanbul Technical University, 34469 Maslak, \.{I}stanbul, Turkey}



\begin{abstract}
	
A combination of a finite number of linear independent states forms superposition in a way that cannot be conceived classically.
Here, using the tools of resource theory of superposition, we give the conditions for a class of superposition state transformations.
These conditions strictly depend on the scalar products of the basis states and reduce to the well-known majorization condition for
quantum coherence in the limit of orthonormal basis. To further superposition-free transformations of $d$-dimensional systems, we
provide superposition-free operators for a deterministic transformation of superposition states. The linear independence of a finite
number of basis states requires a relation between the scalar products of these states. With this information in hand, we determine
the maximal superposition states which are valid over a certain range of scalar products. Notably, we show that, for $d\geq3$, scalar
products of the pure superposition-free states have a greater place in seeking maximally resourceful states.
Various explicit examples illustrate our findings.

\end{abstract}

\maketitle


\section{Introduction}

Research and understanding of the principles of quantum mechanics offer tremendous potential for developing new (quantum) technologies. Quantum superposition \cite{Dirac-Superposition} is one of the most pivotal nonclassical features that can be dealt with in this context. The existence of superposition as a resource delivers significant performance gains on many information processing tasks that cannot be classically achievable. Two particular examples are communication complexity \cite{Feix-CComp, Brukner-Complexity} and the channel discrimination task \cite{Plenio-RTofS}. While the role of superposition in such scenarios is invaluable, it is essential to acquire a thorough understanding of superposition as a resource.

What we know about any study of quantum resource theories (QRTs) \cite{Chitambar-QRTs, Regula-QRTs} is to start with defining
two main elements: the free states and the free operations. The key point is that the free operations must transform free states into free states and allow for the resource to be manipulated but not freely created. From the perspective of QRTs, a leading guide is quantum entanglement \cite{Horodecki-QE}. In the case of entanglement theory, the free states and the free operations are the separable states and local operations and classical communication, respectively. All states which are not free contain resources and are considered costly, and free operations are physical transformations which do not create any resources (for reviews, see Refs.~\cite{Chitambar-QRTs, Regula-QRTs}). Recently, researchers \cite{Sparaciari2020firstlawofgeneral} have extended the tools of QRTs to scenarios in which multiple resources are present and derived conditions for the interconversion of these resources. In Ref.~\cite{Sparaciari2020firstlawofgeneral}, their construction of multiresource theories is based on the definition of their class of allowed operations. Over the past decade a considerable amount of literature has been published on various resource theories \cite{Horodecki-QRT, Veitch_2014, rio2015resource, Brandao-QRT-Reversible, de_Vicente_2016,RT-Heat-and-Work,
Gour-QRT-SingleShot, Liu-ResourceDestroyingMaps, Howard-RTMagic, Gaussin-QRT, Takagi_OpAdvQR, Oszmaniec2019operational, liu2019resource, Wang-RTAsymetric, Li-QuantifyingRCQC, CRT-Incompatibility, kristjnsson2019resource, RT-of-QC, Wolfe2020quantifyingbell}.

The existing body of research on quantum superposition aims to characterize it in all its parts. To this end, quantitative understandings of coherent superposition of quantum states have been achieved \cite{Aberg-Superposition, Plenio-RTofS, Das-NonorthogonalCoherence}.
By relaxing the orthogonality of the basis states to linear independence, superposition theory can be formed as a generalization
of the coherence theory \cite{Baumgratz-Coherence}.
In this sense, Theurer {\textit{et al}}. \cite{Plenio-RTofS} introduced a resource theory of superposition.
There, using the tools of QRTs, superposition-free states and operations were defined, and several quantitative superposition measures were proposed. {\AA}berg \cite{Aberg-Superposition} introduced the concept of superposition measures with respect to given orthogonal decompositions of the Hilbert space of a quantum system.
The notion of quantum coherence for superpositions over states which are not necessarily mutually orthogonal was presented in Ref.~\cite{Das-NonorthogonalCoherence}. Beyond states, coherent superpositions are also possible among quantum evolutions \cite{Bera-QS}.
The author of Ref.~\cite{Bera-QS} developed a resource theoretic framework to quantify superposition present in a quantum evolution.

Given a particular set of resources, a fundamental aspect of any QRT is the manipulation of these resources.
This task deals with whether it is possible to transform one resource into another under free operations.
Several different kinds of manipulation protocols have been studied for certain resources to reveal the importance of resource manipulations. Some of the early examples of research in this area can be found in Refs.~\cite{Nielsen-Maj, AWinter_2004, Bravyi-MagicStatePuri, Brandao-EntManip, Marvian_2013, Vicente_EntManipulation, Wang_AsymptoticEntM, Du-CoherenceMaj, Liu-OneShotRT, Regula-OneShotDistill, Torun-CoherenceDistill, Regula_DepCovOper, Fang-EntDistill, Litinski2019magicstate, Zhao-CoherenceDistill, Chitambar_entManipulation, Fang_No-Go} (see also Refs.~\cite{Horodecki-QE, Chitambar-QRTs, Regula-QRTs, Modi_2012, Streltsov-CoherenceRT, GOUR20151, Goold_2016, Adesso_2016} and papers cited therein). In superposition theory, an initial state can be probabilistically transformed to another target state via superposition-free operations only when the target has an equal or lower superposition rank with the initial state  \cite{Plenio-RTofS}. Nevertheless, under what condition(s) a superposition-free transformation can be achieved deterministically is still an open problem.

In this paper, inspired by Ref.~\cite{Torun-DetCoherence}, we study the transformations of single copies of the pure superposition states in the one-shot setting. More precisely, we provide superposition-free operators for a deterministic transformation and give the conditions for a class of superposition state transformations whereby the tools of resource theory of superposition \cite{Plenio-RTofS} are utilized. These conditions strictly depend on the scalar products of the basis states and reduce to the well-known majorization condition for coherence theory \cite{Du-CoherenceMaj} in the limit of orthonormal basis. Moreover, as a particularly important subject, we determine the maximal superposition states---the state with the greatest resource value---which are valid over a certain range of scalar products. We show that a state with the symmetric superposition of the basis states is the maximally resourceful one for a given set of pure superposition states.
Such contributions are undoubtedly important for our understanding of resource theory of superposition.
To reinforce our findings, we give various examples.

Often researchers are pursuing varied strategies to discover possible connections between different quantum resources. In particular, resource theory of superposition provides insights for the concept of nonclassicality in a way that there can be a conversion between resource states and entanglement, such as faithful transformations where the map transforms free states to separable states---free states for the entanglement theory---and resource states into entangled states \cite{Plenio-Nonclassicality,Plenio-RTofS}. In this context, the resource theory of superposition not only generalizes the resource theory of coherence but also completes the previous work on the notions of nonclassicality and convertibility by adding that linear independence of a set of states is a necessary and sufficient condition for the existence of a faithful transformation \cite{Plenio-RTofS}. In this form, tools introduced in the resource theory of superposition also appear to be a candidate for the quantification of coherence in the optical coherence since optical coherent states are not orthogonal. However, as stated in Ref.~\cite{KCTan2017}, the overcompleteness of the coherent states breaks the condition of linear independence. Our goal is to examine the state transformation problem for resource theory of superposition, where a combination of a finite number of linear independent states forms superposition and the linear independent basis states are pure superposition-free states. Also, to investigate the same problem for linearly dependent basis states our results could be helpful. From the above, a complete study of state transformations is crucial in every sense for the development of resource theories, where our work could find applications in the general theory of nonclassicality.

The structure of the paper is as follows.
Section \ref{Sec:Overview} reviews the basics of the resource theory of superposition.
In Sec.~\ref{Sec:Superposition-freeTr} we discuss the superposition-free transformations.
The linear independence of basis states is truly at the core of the superposition state transformations.
We discuss this crucial point in Sec.~\ref{Subsec:LinearIndependence}.
We then present a clear explanation for the deterministic transformation of superposition states in Sec.~\ref{Subsec:Deterministic}.
We focus on the maximal resourceful states in Sec.~\ref{Sec:Maximal}: qubit systems in Sec.~\ref{Subsec:Max-qubit}
and $d$-dimensional systems in Sec.~\ref{Subsec:Max-d}.
We conclude our work in Sec.~\ref{Sec:Conc}.


\section{Overview of theory}\label{Sec:Overview}

Before scrutinizing the superposition-free transformations, it is useful to review some of the basics of the resource theory of
superposition. For a rigorous resource theory framework for the quantification of superposition we refer to Refs.~\cite{Aberg-Superposition,Plenio-RTofS}.

Let $\{\ket{c_i}\}_{i=1}^{d}$ be a normalized, linear independent and not necessarily orthogonal basis of the Hilbert
space represented by $\mathds{C}^{d}$, $d \in \mathds{N}$.
Any density operator written as
\begin{eqnarray}
\rho=\sum_{i=1}^{d} \rho_{i} \ket{c_i}\bra{c_i},
\end{eqnarray}
where the $\rho_i$ form a probability distribution (i.e., $\rho_i\geq 0$), is called superposition-free. The set of superposition-free density operators is denoted by $\mathcal{F}$ and forms the set of free states. All density operators which are not superposition-free are called superposition states and form the set of resource states~\cite{Plenio-RTofS}. The search problem is to construct a framework for a deterministic transformation of pure superposition states which are given by $\ket{\psi}=\sum_{i=1}^{d}\psi_i\ket{c_i}$.

A Kraus operator $K_i$ is called superposition-free if $K_i \rho K_i^{\dagger}$ $\in$ $\mathcal{F}$  for all $\rho \in \mathcal{F}$. More precisely, a Kraus operator $K_n$ is superposition-free if and only if it is of the form
\begin{eqnarray}\label{Kraus-free-n}
K_n=\sum_{k} c_{k,n} {\ket{c_{f_{n}(k)}}\bra{c_k^{\perp}}},
\end{eqnarray}
where $c_{k,n}\in \mathds{C}$, ${f_{n}(k)}$ are arbitrary index functions~\cite{Plenio-RTofS}, and $\braket{c_i^{\perp}}{c_j}=\zeta_i\delta_{ij}$ for $\zeta_i\in \mathds{C}$ where the vectors $\ket{c_k^{\perp}}$ are normalized. Moreover, quantum operations $\Phi(\rho)$ are called superposition-free if they are trace preserving and can be written such that
$\Phi(\rho)$ $=$ $\sum_{i}K_i \rho K_i^{\dagger}$, where all $K_i$ are free. The set of superposition-free operations
forms the free operations and is denoted by $\mathcal{ FO}$.

One of the most common procedures for defining an order relation between the resource states is related to the concept of free operations. If a state $\rho$ can be transformed into another state $\sigma$ by some free operation, then $\rho$ cannot be less resourceful than $\sigma$ since any task achievable by $\sigma$ is also achievable by $\rho$. However, the converse is not necessarily true. Furthermore, one can introduce resource quantifiers as functionals that preserve this order.
To this goal, the $l_1$ norm of superposition~\cite{Plenio-RTofS} was introduced, and is given by
\begin{eqnarray}
{l_1}(\rho)=\sum_{i\neq j}|\rho_{ij}|,
\end{eqnarray}
for $\rho=\sum_{i,j} \rho_{ij}\ket{c_i}\bra{c_j}$.
We will use the $l_1$ norm of superposition when comparing the resource value of two states.
With these definitions at hand, we are ready to present our protocol and results for the superposition-free transformations.


\section{Superposition-free Transformations}\label{Sec:Superposition-freeTr}

\subsection{Gram Matrix and Linear Independence of Basis States}\label{Subsec:LinearIndependence}

Two particularly important points are worth highlighting.
First, scalar products of the basis states determine the whole structure of the superposition state transformations.
In Sec.~\eqref{Subsec:Deterministic} we will give the conditions for a deterministic transformation that clearly depend on the scalar products. Second, for the linear independence of the basis states $\{\ket{c_i}\}_{i=1}^{d}$, scalar products
must obey a certain inequality.

The {\textit{Gram matrix}} is a useful tool to compute whether a given set
of vectors is linearly independent \cite{AWinter_2004, Marvian_2013, Plenio-Nonclassicality,Regula_2018}.
A set of vectors is linearly independent if and only if the determinant
of the {\textit{Gram matrix}} is positive \cite{Horn-GramMatrix}. Given a finite set of vectors $\{v_1, v_2, \dots, v_m\}$ in an inner products space, the {\textit{Gram matrix}} of the vectors $\{v_1, v_2, \dots, v_m\}$ with respect to the inner product $\braket{\cdot}{\cdot}$ is $G=[\braket{v_n}{v_l}]_{n,l=1,\dots,m} \in M_m$ where $M_m$ is a $m \times m$ square matrix \cite{Horn-GramMatrix}. The Gram matrix is positive definite if and only if the vectors $\{v_1, v_2, \dots, v_m\}$  are linearly independent. Otherwise, it is positive semidefinite. For instance, if one transforms an orthonormal basis to a linear independent basis with a transformation matrix $V$ in a way that $V\ket{i} = \ket{c_i}$, then the Gram matrix equals $V^{\dagger} V$. Moreover, for given two vectors $\ket{\psi} = \sum_{i} \psi_i \ket{c_i}$ and $\ket{\varphi} = \sum_{i} \varphi_i \ket{c_i}$, the inner product can be expressed as $\braket{\psi}{\varphi}=\sum_{ij} G_{ij}\psi_i^{\ast}\varphi_j$ (i.e., the Gram matrix is a metric tensor \cite{Genoni_2019}) where $\psi_i^{\ast}$ is a complex conjugate of $\psi_i$. We show that majorization conditions obtained for the superposition and the coherence theories are related by the Gram matrix (see Appendix \eqref{Sec:App-MatrixApr}).

To obtain the inequality between scalar products, one can construct the corresponding Gram matrix for a given set of basis vectors $\{\ket{c_1}, \ket{c_2}, \dots, \ket{c_d}\}$. Defining $\braket{c_i}{c_j} \coloneqq \mu_{ij}$, then the Gram matrix can be written in the following way:
\begin{eqnarray}\label{gram-matrix}
    G
    =
    \begin{pmatrix}
        1               & \mu_{12}            & \dots  & \mu_{1d} \\
        \mu_{12}^{\ast} & 1                   & \dots  & \mu_{2d} \\
        \vdots          & \vdots              & \ddots & \vdots   \\
        \mu_{1d}^{\ast} & \mu_{2d}^{\ast}     & \dots  & 1        \\

     \end{pmatrix}.
\end{eqnarray}
Consider the case $d = 2$; $ \text{det}(G) = 1 - |\mu_{12}|^2 > 0 $. If we take $\mu_{12} \in \mathds{R}$ it is obvious that $\mu_{12} \in (-1,1)$. However, for $d\geq 3$ the scalar products are constrained with a certain inequality. Considering the case $d=3$ and $\mu_{ij} \in \mathds{R}$, we have
\begin{eqnarray}\label{linear-ind-d3}
    \text{det}(G) = {1-\mu_{12}^2-\mu_{13}^2-\mu_{23}^2+2\mu_{12}\mu_{13}\mu_{23} > 0}.
\end{eqnarray}
Therefore, linear independence of the basis states requires a relation between the scalar products of basis states for $d\geq 3$, e.g., Eq.~\eqref{linear-ind-d3} for $d=3$.

Importantly, since the difficulty of the superposition state transformations is mainly caused by nonorthogonality, we take all the scalar products real and equal throughout the rest of the paper for simplicity and convenience: $\braket{c_i}{c_j}=\mu$ for $i\neq j$. Taking the scalar products as such, one can immediately obtain $-1/2<\mu<1$ for $d=3$ from Eq.~\eqref{linear-ind-d3}. By considering a linear independent set $\{\ket{c_i}\}_{i=1}^{d}$, one can then obtain $1/(1-d)<\mu<1$ for $d\geq 2$. This is one of our starting points to explore superposition-free transformations. We remark that the conditions for a deterministic transformation presented below also hold in nonequal scalar products settings,
i.e., when $\braket{c_i}{c_j}=\mu_{ij}$ for $i\neq j$.


\subsection{Deterministic Transformations of Superposition States}\label{Subsec:Deterministic}

In this section, we present a clear explanation for the deterministic transformation of superposition states.
We consider the transformations between single copies of pure states.
The problem is to transform an initial state $\ket{\psi}$ into a final state $\ket{\varphi}$ under superposition-free operators:
\begin{eqnarray}\label{The-transformation}
\ket{\psi}=\sum_{i=1}^{d}\psi_i \ket{c_i} \overset{\mathcal{FO}}{\longrightarrow}
\ket{\varphi}=\sum_{i=1}^{d}\varphi_i \ket{c_i},
\end{eqnarray}
where the coefficients $\psi_i$ and $\varphi_i$ are real. For the resource theory of superposition it is known that an initial state can be probabilistically (i.e., with some probability $p>0$) transformed to another target state via superposition-free operations only when the target has an equal or lower superposition rank \cite{Plenio-RTofS}. This is also the case for deterministic transformations. Thus, for the superposition states under consideration here, we have $r_S(\ket{\psi}) \geq r_S(\ket{\varphi})$, i.e., the number of nonzero coefficients of $\ket{\psi}$ is equal to or greater than the number of nonzero coefficients of $\ket{\varphi}$.
Since we take all the scalar products equal, the coefficients $\psi_i$ and $\varphi_i$
can be ordered with superposition-free flip operators in a way that
$|\psi_l| \geq |\psi_{l+1}|$ and $|\varphi_l| \geq |\varphi_{l+1}|$ for any $l\in [1,d-1]$.
We also note that the states $\ket{\psi}$ and $\ket{\varphi}$
given in Eq.~\eqref{The-transformation} are normalized, that is $\sum_{i,j}\psi_i(\psi_i+\mu\psi_j)=1$ and
$\sum_{i,j}\varphi_i(\varphi_i+\mu\varphi_j)=1$ where $i\neq j$.

Now we construct superposition-free operators for the transformation given by Eq.~\eqref{The-transformation}.
There are $d!$ different ordering index functions $f_{n}(k)$. Thus, the number of Kraus operators
which leaves the superposition rank of a $d$-dimensional initial state invariant is equal to $d!$ in general \cite{Plenio-RTofS}.
However, in our framework $d$ superposition-free operators are sufficient, and they are given by
\begin{eqnarray}\label{Kraus-d-level1}
K_n=\sum_{k=1}^{d} c_{k,n} \frac{\ket{c_{f_{n}(k)}}\bra{c_k^{\perp}}}{\braket{c_k^{\perp}}{c_k}},
\end{eqnarray}
for $n=1,2,\dots,d$ where $c_{k,n}=\sqrt{p_n}({\varphi_{f_n(k)}}/{\psi_k})$.
The Kraus operators $\{K_n\}_{n=1}^{d}$  given by Eq.~\eqref{Kraus-d-level1} are the most general superposition-free operators which give the desired state with probabilities $\{p_n\}_{n=1}^{d}$, respectively, i.e.,
\begin{eqnarray}\label{Kn-outputs-d-dim}
K_n\ket{\psi}=\sqrt{p_n}\sum_{k=1}^{d}\varphi_{f_n(k)}\ket{c_{f_{n}(k)}}=\sqrt{p_n}\ket{\varphi}.
\end{eqnarray}
where $p_n\geq 0$ and $\sum_{n=1}^{d} p_n=1$. To satisfy the completeness relation,
we introduce another set of superposition-free Kraus operators which are given by \cite{Plenio-RTofS}
\begin{eqnarray}\label{Kraus-d-level2}
F_m=\sum_{k=1}^{d} c_{k,m} \frac{\ket{c_{f_{m}(k)}}\bra{c_k^{\perp}}}{\braket{c_k^{\perp}}{c_k}},
\end{eqnarray}
for $m=(d+1)$, $(d+2)$, $\dots$, $2d$ where $c_{k,m}$ $\in$ $\mathds{C}$. Then the completeness relation is written as
\begin{eqnarray}\label{Completeness-KnFm}
\sum_{n=1}^{d} K_{n}^{\dagger} K_{n}+\sum_{m=d+1}^{2d} F_{m}^{\dagger} F_{m}=I.
\end{eqnarray}
While the Kraus operators defined by Eq.~\eqref{Kraus-d-level1} give the target state [as seen from Eq.~\eqref{Kn-outputs-d-dim}],
the Kraus operators $\{F_m\}_{m=(d+1)}^{2d}$ give nothing, i.e., $F_m\ket{\psi}=0$.

\begin{table}[t]
\caption{The table shows us the order of the index functions $f_n(k)$. From \eqref{Kn-outputs-d-dim} we have outputs $\sum_{k=1}^{d}\varphi_{f_n(k)}\ket{c_{f_{n}(k)}}$, and here we give the values of $\{f_n(k)\}_{k=1}^{d}$ for $d=2,3,4$ and $n=1,2,\dots,d$: The first row of the table shows the index functions for the Kraus operators $K_1$ and $K_2$, respectively;
The second and third rows of the table correspond to a particular case of $d=3$, $\tilde{\psi}_2 \geq \tilde{\varphi}_2$ for
the former and $\tilde{\psi}_2 \leq \tilde{\varphi}_2$ for the latter; and the rest are for four-dimensional systems where each one corresponds to a particular case, e.g., the fourth row of the table (first case of $d=4$) corresponds to the case $\tilde{\psi}_2 \geq \tilde{\varphi}_2$ and $\tilde{\psi}_3 \geq \tilde{\varphi}_3$. 
}
\label{Table:Permutations-d234}
\centering
\begin{ruledtabular}
\begin{tabular}{c c c c c}
                       & \multicolumn{4}{c}{Index functions $\{f_n(k)\}_{k=1}^{d}$}       \Tstrut\Bstrut \\ [1ex] \cline{2-5}
                       & $\{f_1(k)\}_{k=1}^{d}$  & $\{f_2(k)\}_{k=1}^{d}$    & $\{f_3(k)\}_{k=1}^{d}$   & $\{f_4(k)\}_{k=1}^{d}$      \Tstrut\Bstrut \\ [1ex] \hline
\multirow{1}{*}{$d=2$} & $\{1,2\}$     & $\{2,1\}$       & -              & -                 \Tstrut \\ [1ex]
\multirow{2}{*}{$d=3$} & $\{1,2,3\}$   & $\{3,2,1\}$     & $\{2,1,3\}$    & -                 \\
                       & $\{1,2,3\}$   & $\{3,2,1\}$     & $\{1,3,2\}$    & -                 \\ [1ex]
\multirow{5}{*}{$d=4$} & $\{1,2,3,4\}$ & $\{4,2,3,1\}$   & $\{2,1,3,4\}$  & $\{3,2,1,4\}$     \\
                       & $\{1,2,3,4\}$ & $\{4,2,3,1\}$   & $\{2,1,3,4\}$  & $\{1,2,4,3\}$     \\
                       & $\{1,2,3,4\}$ & $\{4,2,3,1\}$   & $\{1,4,3,2\}$  & $\{1,2,4,3\}$     \\
                       & $\{1,2,3,4\}$ & $\{4,2,3,1\}$   & $\{1,4,3,2\}$  & $\{1,3,2,4\}$     \\
                       & $\{1,2,3,4\}$ & $\{4,2,3,1\}$   & $\{3,2,1,4\}$  & $\{1,3,2,4\}$     \\
\end{tabular}
\end{ruledtabular}
\end{table}

The next step is to determine the index functions $f_{n}(k)$. To this goal, we benefit from the results of deterministic transformations of coherent states under incoherent operations presented in Ref.~\cite{Torun-DetCoherence}.
There, incoherent Kraus operators were constructed for $d$-dimensional systems by explicitly presenting permutations. These permutations provide us the index functions $f_{n}(k)$. We then define
\begin{eqnarray}
\ket{c_{f_{1}(k)}} & \coloneqq & \ket{c_{k}} , \\
\ket{c_{f_{m}(k)}} & \coloneqq & \ket{c_{m-d}},
\end{eqnarray}
and for each $n \in [2,d]$ there is a pair $(\alpha, \beta) \in [1,d]$ such that
\begin{eqnarray}\label{alpha-beta-permutation}
\ket{c_{f_{n}(\alpha)}}  \coloneqq  \ket{c_{\beta}}, \quad
\ket{c_{f_{n}(\beta)}} \coloneqq  \ket{c_{\alpha}}.
\end{eqnarray}
This corresponds to the permutation $\ket{\alpha}$ $\leftrightarrow$ $\ket{\beta}$ for coherence transformations~\cite{Torun-DetCoherence}. Then we have
\begin{eqnarray}
\ket{c_{f_{n}(\gamma)}}  \coloneqq  \ket{c_{\gamma}},
\end{eqnarray}
for $n \in [2,d]$ and $\gamma=1,2,\dots,d$ but $\gamma \neq \alpha$ and $\gamma \neq \beta$.
To clarify the above definitions which are related with the results \cite{Torun-DetCoherence} we give the terms $\{f_n(k)\}_{k=1}^{d}$ for $d=2,3,4$ (see Table \eqref{Table:Permutations-d234}).


So far, we have introduced the superposition-free Kraus operators to be used. Now we investigate the condition(s) for superposition-free transformations. The authors of Ref.~\cite{Du-CoherenceMaj} built the counterpart of the celebrated Nielsen theorem \cite{Nielsen-Maj} for coherence manipulations and showed that majorization is the necessary and sufficient condition for a deterministic transformation. In this respect, in principle, a similar approach is highly expected for superposition manipulation. In the following we give the condition(s) for the superposition-free transformations given by Eq.~\eqref{The-transformation}.

The completeness relation given by Eq.~\eqref{Completeness-KnFm} is essential for us to investigate condition(s) for deterministic transformations. We start by defining
\begin{eqnarray}\label{Tilde-psi-varphi}
\tilde{\psi}_i \coloneqq \psi_i(\psi_i+\mu\mathop{\sum_{j=1}^{d}}_{(j\neq i)}\psi_j),
\quad
\tilde{\varphi}_i \coloneqq \varphi_i(\varphi_i+\mu\mathop{\sum_{j=1}^{d}}_{(j\neq i)}\varphi_j),
\end{eqnarray}
for $i=1,2,\dots,d$ where the coefficients $\tilde{\psi}_i$ and $\tilde{\varphi}_i$ are in an order such that
$\tilde{\psi}_l\geq\tilde{\psi}_{l+1}$ and $\tilde{\varphi}_l\geq\tilde{\varphi}_{l+1}$ for any $l\in [1,d-1]$.
Using superposition-free operators given by Eqs.~\eqref{Kraus-d-level1} and \eqref{Kraus-d-level2} it may be possible to transform
$\ket{\psi}$ into another state $\ket{\varphi}$ deterministically if the majorization condition (for superposition theory) is satisfied:
\begin{eqnarray}\label{Maj-Superposition-Condition1}
{\text{Majorization}}: \ \ \sum_{i=1}^{k}\tilde{\psi}_i \leq \sum_{i=1}^{k}\tilde{\varphi}_i,
\end{eqnarray}
for any $k \in [1,d]$ where equality holds for $k=d$. Contrary to coherence manipulation \cite{Du-CoherenceMaj}, majorization alone is not the necessary and sufficient condition for superposition-free transformations. The completeness equation \eqref{Completeness-KnFm} also dictates one more condition to be satisfied, which we call the condition on completeness (CoC). It is given such that
\begin{eqnarray}\label{Condition2}
{\text{CoC}}: \ \ \sum_{i=1}^{d} p_i \omega_{ij} \leq \psi_j^2,
\end{eqnarray}
for $j=2, 3, \dots, d$. Here $\omega_{ij}$ is the $(ij)$th element of a $d\times d$ matrix $\omega$.
There exists a permutation matrix $P_i$ such that
\begin{eqnarray}\label{CoC-wij}
\left(\begin{array}{c}  \omega_{i1} \\ \omega_{i2}  \\ \vdots \\ \omega_{id} \end{array}\right)
= P_i  \left(\begin{array}{c}  \varphi_1^2 \\ \varphi_2^2  \\ \vdots \\ \varphi_d^2 \end{array}\right).
\end{eqnarray}
This $P_i$ corresponds with the permutation $\ket{\alpha}$ $\leftrightarrow$ $\ket{\beta}$ given in
Eq.~\eqref{alpha-beta-permutation}. Also, the first row of the matrix $\omega$ is equal to
$(\varphi_1^2, \varphi_2^2, \dots, \varphi_d^2)$; i.e., $P_1$ is identity. To obtain the probabilities $\{p_n\}_{n=1}^{d}$ in Eq.~\eqref{Kn-outputs-d-dim} [and in Eq.~\eqref{Condition2}] one needs to solve the following equations:
\begin{eqnarray}\label{probabilities-d-dim}
\sum_{n=1}^{d} p_n \tilde{\varphi}_{f_n(k)} = \tilde{\psi}_k, \quad k=1,2,\dots,d.
\end{eqnarray}

\begin{figure}[t]
	\centering
	\includegraphics[width=1\columnwidth]{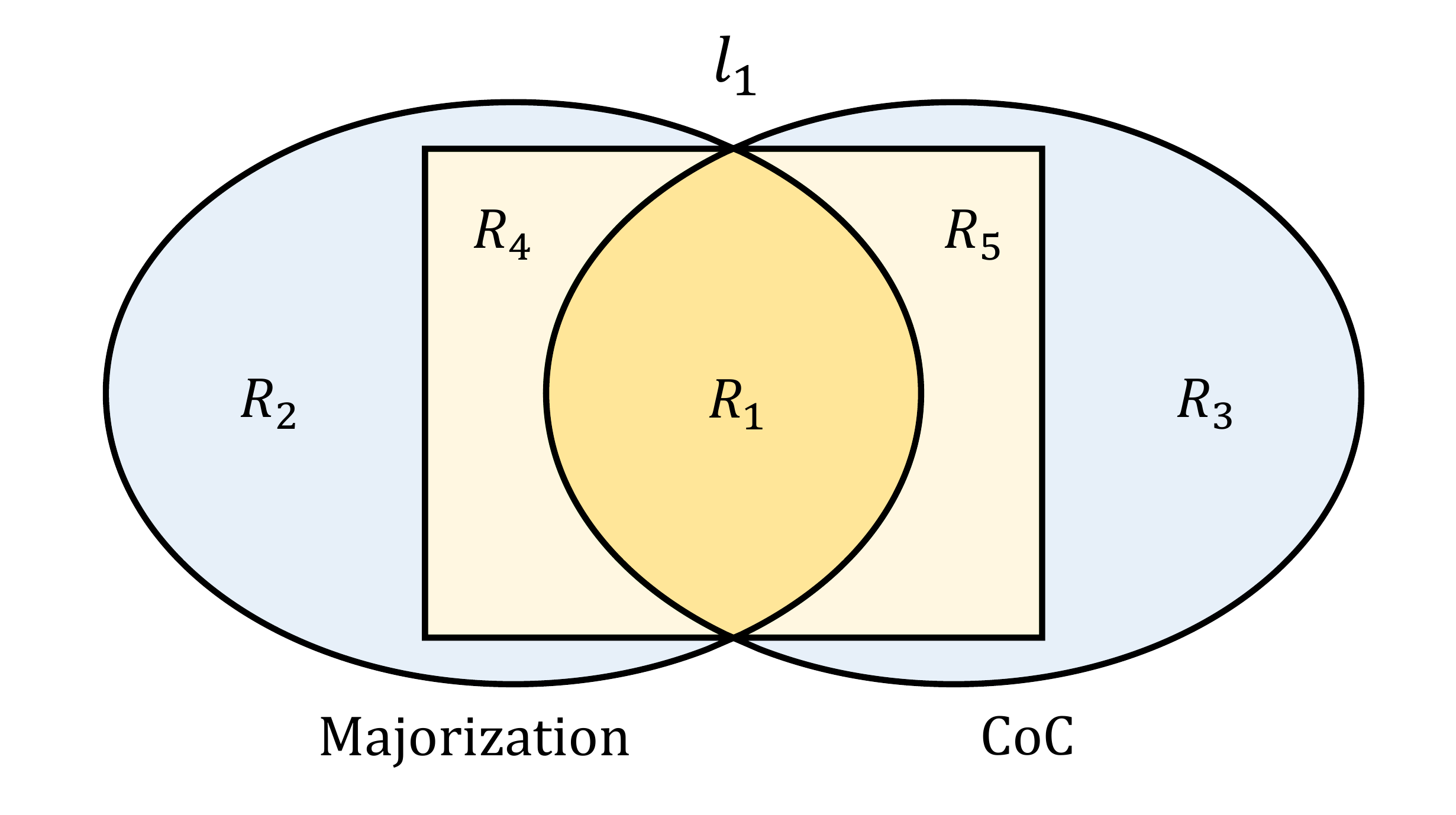}
	\caption{An illustrative diagram for superposition-free transformations. For a given initial state $\ket{\psi}$ and final state $\ket{\varphi}$, each region $R_i$, $i=1,\dots,5$, shows whether it provides majorization (for superposition), condition on completeness (CoC), and $l_1$ norm. In the region $R_2$ the initial state $\ket{\psi}$ and the target state $\ket{\varphi}$ satisfy majorization and do not satisfy the $l_1$ norm and CoC; in the region $R_4$ the initial and target states satisfy the $l_1$ norm and majorization but do not satisfy CoC, and similarly for other regions. Moreover, the region $R_1$ gives the conditions for a class of deterministic transformation of superposition states: For a deterministic transformation, the initial and the target states must satisfy the CoC and majorization conditions, and the $l_1$ norm of superposition [by this we mean $l_1(\ket{\psi}) \geq l_1(\ket{\varphi})$]. Explicit examples are given in the paper.}
	\label{figure1}
\end{figure}

When Eqs.~\eqref{Maj-Superposition-Condition1} and \eqref{Condition2} are both satisfied for a transformation $\ket{\psi}\overset{\mathcal{FO}}{\longrightarrow}\ket{\varphi}$, then $l_1(\ket{\psi}\bra{\psi}) \geq l_1(\ket{\varphi}\bra{\varphi})$ (see Fig.~\eqref{figure1}); i.e., neither is sufficient alone for a deterministic transformation. In addition, the theory of superposition contains coherence theory as a special case. In this respect, when the basis states are orthogonal, i.e., for $\mu=0$, Eq.~\eqref{Maj-Superposition-Condition1} turns into the well known majorization condition for coherence \cite{Du-CoherenceMaj} and the equality holds in Eq.~\eqref{Condition2} as well.
Obtaining Eqs.~\eqref{Maj-Superposition-Condition1} and \eqref{Condition2} requires some algebra  which we do in Appendix \eqref{Sec:App-ConditionsProof}.



\subsubsection{Qubit systems}

To establish a useful and efficient protocol for the superposition-free transformation, first the problem
should be solved in all its details for the simplest case, $d=2$. In this direction, let us consider
the transformation
\begin{eqnarray}\label{The-transformation-qubit}
\psi_1  e^{i\alpha_1} \ket{c_1}+\psi_2 e^{i\alpha_2} \ket{c_2} \overset{\mathcal{FO}}{\longrightarrow}
\varphi_1 e^{i\beta_1} \ket{c_1}+\varphi_2 e^{i\beta_2} \ket{c_2},
\end{eqnarray}
where $\alpha_i \in [0,2\pi]$, $\beta_i \in [0,2\pi]$, and $\braket{c_1}{c_2}=\mu$.
This is the most general transformation for qubit systems.
The states are normalized as usual and we can choose, without loss of generality, $\alpha_1=\beta_1=0$.
In what follows we show that different choices of local phases $\alpha_2$ and $\beta_2$ produce different results.

We start with the case $\alpha_2=\beta_2=0$.
There is no local phase for both the initial and the final states.
By using the superposition-free operators
\begin{eqnarray}\label{K1-qubit}
K_1=\sqrt{p_1}\Big(\frac{\varphi_1}{\psi_1}\frac{\ket{c_1}\bra{c_1^{\perp}}}{\braket{c_1^{\perp}}{c_1}}
+\frac{\varphi_2}{\psi_2}\frac{\ket{c_2}\bra{c_2^{\perp}}}{\braket{c_2^{\perp}}{c_2}}\Big),
\end{eqnarray}
\begin{eqnarray}\label{K2-qubit}
K_2=\sqrt{p_2}\Big(\frac{\varphi_2}{\psi_1}\frac{\ket{c_2}\bra{c_1^{\perp}}}{\braket{c_1^{\perp}}{c_1}}
+\frac{\varphi_1}{\psi_2}\frac{\ket{c_1}\bra{c_2^{\perp}}}{\braket{c_2^{\perp}}{c_2}}\Big),
\end{eqnarray}
$F_3$, and $F_4$ defined by Eq.~\eqref{Kraus-d-level2},
one can achieve the transformation $\ket{\psi} \overset{\mathcal{FO}}{\longrightarrow}\ket{\varphi}$ deterministically where $K_1\ket{\psi}=\sqrt{p_1}\ket{\varphi}$, $K_2\ket{\psi}=\sqrt{p_2}\ket{\varphi}$, $F_3\ket{\psi}=0$, and $F_4\ket{\psi}=0$.
The completeness condition defined by Eq.~\eqref{Completeness-KnFm}
gives us three equations. By using Eq.~\eqref{Tilde-psi-varphi}, the first two can be written as
\begin{eqnarray}
p_1\tilde{\varphi}_1+p_2\tilde{\varphi}_2=\tilde{\psi}_1, \quad
p_1\tilde{\varphi}_2+p_2\tilde{\varphi}_1=\tilde{\psi}_2,
\end{eqnarray}
where $p_1+p_2=1$. Then the probabilities are found to be
\begin{eqnarray}\label{2D-probabilites}
p_1=\frac{\tilde{\varphi}_1-\tilde{\psi}_2}{\tilde{\varphi}_1-\tilde{\varphi}_2}, \quad p_2=\frac{\tilde{\psi}_2-\tilde{\varphi}_2}{\tilde{\varphi}_1-\tilde{\varphi}_2}.
\end{eqnarray}
The positivity of the probabilities given by Eq.~\eqref{2D-probabilites} leads to the condition \eqref{Maj-Superposition-Condition1}. Another constraint on the Kraus operators comes from the third equation of the completeness condition \eqref{Completeness-KnFm} which implies
\begin{eqnarray}\label{F3F4-qubit-explicit}
c_{2,3}^2+c_{2,4}^2=\left(\frac{\mu}{\psi_2^2}\right)(\varphi_1\varphi_2-\psi_1\psi_2).
\end{eqnarray}
The left-hand side of the above is non-negative, yielding $\mu(\varphi_1\varphi_2-\psi_1\psi_2)\geq 0$.
This inequality can be written such that
\begin{eqnarray}\label{CoC-Qubit-1st}
p_1\varphi_2^2+p_2\varphi_1^2 \leq \psi_2^2,
\end{eqnarray}
which gives us the condition \eqref{Condition2} for qubit systems. Thus, a single condition, Eq.~\eqref{Maj-Superposition-Condition1}, is not sufficient and one more condition, Eq.~\eqref{Condition2}, is necessary.
Also, equality holds in Eq.~\eqref{CoC-Qubit-1st} and the results become the same as in Ref.~\cite{Torun-DetCoherence} in the limit of orthonormal basis.

Note that the explicit construction of $F_3$ and $F_4$ is not necessary \cite{Plenio-RTofS}.
Here, different choices of $c_{2,3}$ and $c_{2,4}$ given in Eq.~\eqref{F3F4-qubit-explicit} give us different sets of $\{F_3, F_4\}$, provided that Eq.~\eqref{F3F4-qubit-explicit} is satisfied. For instance, if we choose $c_{2,4}=0$ (or $c_{2,3}=0$) then we have only three superposition-free Kraus operators, $\{K_1,K_2, F_3\}$ (or $\{K_1,K_2, F_4\}$). On the other hand, for $c_{2,3}\neq 0$ and $c_{2,4}\neq 0$, we need four superposition-free Kraus operators to make the entire operation trace preserving.

To better understand what we aim to depict by Fig.~\eqref{figure1}, let us consider the following examples. We have an initial state $\ket{\psi}$ and a target state $\varphi$; i.e., we have a pair of superposition states such that $\{\ket{\psi},\ket{\varphi}\}$, for each region $R_i$, $i=1,\dots,5$:
\begin{eqnarray}\label{R1}
R_1: \ \ \Big\{\frac{3\ket{c_1}-\ket{c_2}}{\sqrt{7}}, \ \
\frac{4\ket{c_1}-\ket{c_2}}{\sqrt{13}}\Big\},
\end{eqnarray}
\begin{eqnarray}\label{R2}
R_2: \ \ \Big\{\frac{3\ket{c_1}+\ket{c_2}}{\sqrt{13}}, \ \
\frac{4\ket{c_1}-\ket{c_2}}{\sqrt{13}}\Big\},
\end{eqnarray}
\begin{eqnarray}\label{R3}
R_3: \ \ \Big\{\frac{4\ket{c_1}+\ket{c_2}}{\sqrt{21}}, \ \
\frac{3\ket{c_1}+\ket{c_2}}{\sqrt{13}}\Big\},
\end{eqnarray}
\begin{eqnarray}\label{R4}
R_4: \ \ \Big\{\frac{3\ket{c_1}+\ket{c_2}}{\sqrt{13}}, \ \
\frac{4\ket{c_1}+\ket{c_2}}{\sqrt{21}}\Big\},
\end{eqnarray}
\begin{eqnarray}\label{R5}
R_5: \ \ \Big\{\frac{4\ket{c_1}-\ket{c_2}}{\sqrt{13}}, \ \
\frac{3\ket{c_1}+\ket{c_2}}{\sqrt{13}}\Big\},
\end{eqnarray}
and $\mu=1/2$. The majorization condition given by Eq.~\eqref{Maj-Superposition-Condition1} and CoC given by Eq.~\eqref{Condition2} are satisfied only for the pair of states given in Eq.~\eqref{R1}.
Then it is easy to show that the transformation $\ket{\psi}\overset{\mathcal{FO}}{\longrightarrow}\ket{\varphi}$ given in Eq.~\eqref{R1} can be achieved deterministically
by using the superposition-free operators
\begin{eqnarray}\label{K1-qubit-examplee}
K_1=\sqrt{\frac{209}{210}}\Big[\sqrt{\frac{7}{13}}\Big(\frac{4}{3}\frac{\ket{c_1}\bra{c_1^{\perp}}}{\braket{c_1^{\perp}}{c_1}}
+\frac{\ket{c_2}\bra{c_2^{\perp}}}{\braket{c_2^{\perp}}{c_2}}\Big)\Big],
\end{eqnarray}
\begin{eqnarray}\label{K2-qubit-examplee}
K_2=-\sqrt{\frac{1}{210}}\Big[\sqrt{\frac{7}{13}}\Big(\frac{1}{3}\frac{\ket{c_2}\bra{c_1^{\perp}}}{\braket{c_1^{\perp}}{c_1}}
+4\frac{\ket{c_1}\bra{c_2^{\perp}}}{\braket{c_2^{\perp}}{c_2}}\Big)\Big],
\end{eqnarray}
\begin{eqnarray}\label{F3-qubit-examplee}
F_3=-\sqrt{\frac{11}{26}}\Big(\frac{1}{3}\frac{\ket{c_1}\bra{c_1^{\perp}}}{\braket{c_1^{\perp}}{c_1}}
+\frac{\ket{c_1}\bra{c_2^{\perp}}}{\braket{c_2^{\perp}}{c_2}}\Big),
\end{eqnarray}
where $K_1\ket{\psi}=\sqrt{\frac{209}{210}}\ket{\varphi}$, $K_2\ket{\psi}=\frac{1}{\sqrt{210}}\ket{\varphi}$, and $F_3\ket{\psi}=0$.
We here take $c_{2,4}=0$ in Eq.~\eqref{F3F4-qubit-explicit}, and therefore $F_4=0$.
However, for the pairs of states given in Eqs.~\eqref{R2}--\eqref{R5} a deterministic transformation is not possible under superposition-free operations while conditions \eqref{Maj-Superposition-Condition1} and \eqref{Condition2}  are not satisfied at the same time. As it is seen from the given examples, the $l_1$ norm of superposition of the initial state is greater than the final state both for given examples \eqref{R4} and \eqref{R5}. Furthermore, the transformations given in Eqs.~\eqref{R2} and \eqref{R5} and Eqs.~\eqref{R3} and \eqref{R4} are obviously the opposite of each other, i.e., neither $\ket{\psi}\overset{\mathcal{FO}}{\longrightarrow}\ket{\varphi}$ nor
$\ket{\varphi}\overset{\mathcal{FO}}{\longrightarrow}\ket{\psi}$ is a deterministic transformation.
Overall, for a given initial state $\ket{\psi}$ and final state $\ket{\varphi}$, $l_1(\ket{\psi}) \geq l_1(\ket{\varphi})$ does not necessarily mean that the state $\ket{\psi}$ can be transformed into the state $\ket{\varphi}$ with unit probability under the superposition-free operators.

Defining ${\psi_1}/{\psi_2} \coloneqq \lambda$ and ${\varphi_1}/{\varphi_2} \coloneqq \kappa$ and
after some algebra, it is possible to reduce the conditions for a deterministic transformation, for qubit systems,
into the following forms:
\begin{eqnarray}\label{lambda-negative}
\lambda < 0 \ \Rightarrow \ {0 \leq \mu < -\frac{\kappa+\lambda}{1+\kappa\lambda}},
\end{eqnarray}
or
\begin{eqnarray}\label{lambda-positive}
\lambda > 0 \ \Rightarrow \ {-\frac{\kappa+\lambda}{1+\kappa\lambda} < \mu \leq 0},
\end{eqnarray}
where $|\kappa|\geq|\lambda|$ for both cases.
The inferences about Eqs.~\eqref{lambda-negative} and \eqref{lambda-positive} are fairly straightforward.
A deterministic transformation can be achieved only for certain values of scalar product of basis states depending on whether the $\lambda$ is negative or positive. Furthermore, Eqs.~\eqref{lambda-negative} and \eqref{lambda-positive} exhibit a clear observation about maximal superposition states for qubit systems. If $\lambda=-1$ then $0\leq \mu < 1$ and if $\lambda=1$ then $-1 < \mu \leq 0$. Thus, there are two maximal superposition states for qubit systems: one is $\psi_1=-\psi_2$ with $\mu \in [0,1)$ and the other is $\psi_1=\psi_2$ with $\mu \in (-1,0]$. We will discuss the maximal superposition states in Sec.~\eqref{Sec:Maximal}. However, seeking maximal superposition states becomes dramatically harder for $d>2$.

We conclude by briefly considering two other possible cases: only the final state has a local phase, i.e., $\alpha_2=0$ and $\beta_2\neq0$, or only the initial state has a local phase, i.e., $\alpha_2\neq0$ and $\beta_2=0$.
While a deterministic transformation is possible for the former 
only when $\psi_1=\mp\psi_2$, a deterministic transformation is not possible for the latter.
By constructing Kraus operators for the case $\alpha_2\neq0$ and $\beta_2=0$,
one obtains an equation such that $\kappa\lambda\mu \sin\alpha_2=0$.
However, this equation is satisfied only for the orthogonal limit, i.e., for $\mu=0$.
Therefore, in the case of superposition, a deterministic transformation is not possible when only the initial state has a local phase.



\subsubsection{Three-dimensional systems}

Once the deterministic transformation of superposition states have been presented for qubit systems,
we can now systematically examine the same problem for $d=3$. The transformation under investigation
is as follows:
\begin{eqnarray}\label{The-transformation-3d}
\sum_{i=1}^{3}\psi_i \ket{c_i} \overset{\mathcal{FO}}{\longrightarrow}
\sum_{i=1}^{3}\varphi_i \ket{c_i},
\end{eqnarray}
where there is no local phase for both initial and final states.
We take advantage of the results presented in Ref.~\cite{Torun-DetCoherence}.
To this goal, using Eqs.~\eqref{Kraus-d-level1} and \eqref{Kraus-d-level2},
we explicitly give the terms $c_{k,n}=\sqrt{p_n}({\varphi_{f_n(k)}}/{\psi_k})$ and $\ket{c_{f_{n}(k)}}$ for $k=1,2,3$ and $n=1,2,3$, and construct the Kraus operators step by step.
To enhance the understanding of high-dimensional solutions, it is useful to proceed in this way.
For $d=3$, we have three Kraus operators such that
\begin{eqnarray}
K_n\sum_{i=1}^{3}\psi_i \ket{c_i}
&=&\sqrt{p_n}\sum_{k=1}^{3}\varphi_{f_n(k)}\ket{c_{f_{n}(k)}} \nonumber \\
&=&\sqrt{p_n}\sum_{i=1}^{3}\varphi_i \ket{c_i}.
\end{eqnarray}
We stress that the solutions of the three-dimensional systems are divided into two subcases (see Table \eqref{Table:Permutations-d234}).
From Table \eqref{Table:Permutations-d234}, $\{{f_{1}(k)}\}_{k=1}^{3}=\{1,2,3\}$. Then, for the Kraus operator $K_1$ (for both subcases) we have
\begin{eqnarray}\label{D3-probablities-K1}
c_{1,1}&=&\sqrt{p_1} \big({\varphi_1}/{\psi_1}\big), \quad \ket{c_{f_{1}(1)}}=\ket{c_1}, \nonumber \\
c_{2,1}&=&\sqrt{p_1} \big({\varphi_2}/{\psi_2}\big), \quad \ket{c_{f_{1}(2)}}=\ket{c_2}, \nonumber \\
c_{3,1}&=&\sqrt{p_1} \big({\varphi_3}/{\psi_3}\big), \quad \ket{c_{f_{1}(3)}}=\ket{c_3}.
\end{eqnarray}
When we perform Kraus operator $K_1$ to the initial state, we obtain $\varphi_1\ket{c_1}+\varphi_2\ket{c_2}+\varphi_3\ket{c_3}$
with probability $p_1$. 
As seen from Table \eqref{Table:Permutations-d234}  $\{{f_{2}(k)}\}_{k=1}^{3}=\{3,2,1\}$.
Then, for the Kraus operator $K_2$ (for both subcases) we have
\begin{eqnarray}\label{D3-probablities-K2}
c_{1,2}&=&\sqrt{p_2} \big({\varphi_3}/{\psi_1}\big), \quad \ket{c_{f_{2}(1)}}=\ket{c_3}, \nonumber \\
c_{2,2}&=&\sqrt{p_2} \big({\varphi_2}/{\psi_2}\big), \quad \ket{c_{f_{2}(2)}}=\ket{c_2}, \nonumber \\
c_{3,2}&=&\sqrt{p_2} \big({\varphi_1}/{\psi_3}\big), \quad \ket{c_{f_{2}(3)}}=\ket{c_1}.
\end{eqnarray}
This means our $(\alpha, \beta) \in [1,d]$ pair given in Eq.~\eqref{alpha-beta-permutation} is $(1,3)$.
When we perform Kraus operator $K_2$ to the initial state, we obtain $\varphi_3\ket{c_3}+\varphi_2\ket{c_2}+\varphi_1\ket{c_1}$
with probability $p_2$. 

Although the relations $\tilde{\psi}_1 \leq \tilde{\varphi}_1$ and $\tilde{\psi}_3 \geq \tilde{\varphi}_3$ follow from
the majorization conditions given by Eq.~\eqref{Maj-Superposition-Condition1}, there are two possible relations between $\tilde{\psi}_2$ and $\tilde{\varphi}_2$. The operator $K_3$ has two different forms depending on the two different relations between the parameters of the source and target states: the terms $c_{k,n}$ and $\ket{c_{f_{n}(k)}}$ are different for $\tilde{\psi}_2 \geq \tilde{\varphi}_2$ and $\tilde{\psi}_2 \leq \tilde{\varphi}_2$. These two subcases together solve the problem for three-dimensional systems completely. In Table \eqref{Table:Permutations-d234} we give index functions $f_{n}(k)$ for $d=3$: the second row of Table \eqref{Table:Permutations-d234} for $\tilde{\psi}_2 \geq \tilde{\varphi}_2$ and the third row of Table \eqref{Table:Permutations-d234} for $\tilde{\psi}_2 \leq \tilde{\varphi}_2$.
We proceed to solve the problem under these two subcases one by one.

{\emph{The case $\tilde{\psi}_2 \geq \tilde{\varphi}_2$}}:
From the second row of Table \eqref{Table:Permutations-d234} $\{{f_{3}(k)}\}_{k=1}^{3}=\{2,1,3\}$.
Then, the terms $c_{k,n}$ and $\ket{c_{f_{n}(k)}}$ for the Kraus operator $K_3$ are given by
\begin{eqnarray}\label{D3-probablities-first-case}
c_{1,3}&=&\sqrt{p_3} \big({\varphi_2}/{\psi_1}\big), \quad \ket{c_{f_{3}(1)}}=\ket{c_2}, \nonumber \\
c_{2,3}&=&\sqrt{p_3} \big({\varphi_1}/{\psi_2}\big), \quad \ket{c_{f_{3}(2)}}=\ket{c_1}, \nonumber \\
c_{3,3}&=&\sqrt{p_3} \big({\varphi_3}/{\psi_3}\big), \quad \ket{c_{f_{3}(3)}}=\ket{c_3}.
\end{eqnarray}
This means our $(\alpha, \beta) \in [1,d]$ pair given in Eq.~\eqref{alpha-beta-permutation} is $(1,2)$.
When we perform Kraus operator $K_3$ to the initial state, we obtain $\varphi_2\ket{c_2}+\varphi_1\ket{c_1}+\varphi_3\ket{c_3}$
with probability $p_3$.
The Kraus operators $F_4$, $F_5$, and $F_6$ are given by Eq.~\eqref{Kraus-d-level2} in a way that yields $\sum_{n=1}^{3} K_{n}^{\dagger} K_{n}+\sum_{m=4}^{6} F_{m}^{\dagger} F_{m}=I$.
For this case, the probabilities in Eqs.~\eqref{D3-probablities-K1}, \eqref{D3-probablities-K2}, and \eqref{D3-probablities-first-case},
are found to be
\begin{eqnarray}\label{probabilities-3d-first-case}
p_1=1-p_2-p_3, \quad
p_2=\frac{\tilde{\psi}_3-\tilde{\varphi}_3}{\tilde{\varphi}_1-\tilde{\varphi}_3}, \quad
p_3=\frac{\tilde{\psi}_2-\tilde{\varphi}_2}{\tilde{\varphi}_1-\tilde{\varphi}_2}.
\end{eqnarray}
We stress that both the majorization condition \eqref{Maj-Superposition-Condition1} and CoC \eqref{Condition2} need to be satisfied (as seen, the condition for the positivity of the probabilities implies $\tilde{\psi}_3 \geq \tilde{\varphi}_3$ which is just the majorization relation and $\tilde{\psi}_2 \geq \tilde{\varphi}_2$ which is the case we are interested in), and the corresponding $\omega$ matrix is given by
\begin{eqnarray}\begin{aligned}
\omega=\left(\begin{array}{cccccc}  \varphi_1^2 & \varphi_2^2 & \varphi_3^2  \\ \varphi_3^2 & \varphi_2^2 & \varphi_1^2 \\ \varphi_2^2 & \varphi_1^2 & \varphi_3^2 \end{array}\right).
\end{aligned}\end{eqnarray}
Let us consider the following two examples. We have an initial superposition state
\begin{eqnarray}\label{initial-state-3d-firtscase}
\ket{\psi}=\sqrt{\frac{2}{17}}\big({3\ket{c_1}+2\ket{c_2}+\ket{c_3}}\big),
\end{eqnarray}
with $\mu=-1/4$. Then, for a given target state
\begin{eqnarray}\label{final-state1-3d-firtscase}
\ket{\varphi_1}=\frac{1}{{\sqrt{14}}}\big({4\ket{c_1}+2\ket{c_2}+\ket{c_3}}\big),
\end{eqnarray}
the transformation $\ket{\psi}\overset{\mathcal{FO}}{\longrightarrow}\ket{\varphi_1}$ can be achieved deterministically by using the Kraus operators defined above, i.e., superposition-free Kraus operators are given by
\begin{eqnarray}\label{K1-3d-1st}
K_1=\sqrt{p_1}\sqrt{\frac{17}{28}}\Big(\frac{4}{3}\frac{\ket{c_1}\bra{c_1^{\perp}}}{\zeta_1}
+\frac{\ket{c_2}\bra{c_2^{\perp}}}{\zeta_2}
+\frac{\ket{c_3}\bra{c_3^{\perp}}}{\zeta_3}\Big), \quad
\end{eqnarray}
\begin{eqnarray}\label{K2-3d-1st}
K_2=\sqrt{p_2}\sqrt{\frac{17}{28}}\Big(\frac{1}{3}\frac{\ket{c_3}\bra{c_1^{\perp}}}{\zeta_1}
+\frac{\ket{c_2}\bra{c_2^{\perp}}}{\zeta_2}
+4\frac{\ket{c_1}\bra{c_3^{\perp}}}{\zeta_3}\Big), \quad
\end{eqnarray}
\begin{eqnarray}\label{K3-3d-1st}
K_3=\sqrt{p_3}\sqrt{\frac{17}{28}}\Big(\frac{2}{3}\frac{\ket{c_2}\bra{c_1^{\perp}}}{\zeta_1}
+2\frac{\ket{c_1}\bra{c_2^{\perp}}}{\zeta_2}
+\frac{\ket{c_3}\bra{c_3^{\perp}}}{\zeta_3}\Big), \quad
\end{eqnarray}
where $\zeta_i=\braket{c_i^{\perp}}{c_i}$ for $i=1,2,3$, $p_1=\frac{2947}{3519}$, $p_2=\frac{1}{153}$, and $p_3=\frac{61}{391}$.
Also, Kraus operators $F_4$, $F_5$, and $F_6$ are given by Eq.~\eqref{Kraus-d-level2}.
As mentioned before, it is not necessary to obtain these Kraus operators explicitly, provided that the CoC given by Eq.~\eqref{F3F4-qubit-explicit} is satisfied. However, for a given target state
\begin{eqnarray}\label{final-state2-3d-firtscase}
\ket{\varphi_2}=\frac{1}{{2\sqrt{5}}}\big({4\ket{c_1}+2\ket{c_2}-\ket{c_3}}\big),
\end{eqnarray}
the transformation $\ket{\psi}\overset{\mathcal{FO}}{\longrightarrow}\ket{\varphi_2}$ cannot be achieved deterministically although the $l_1$ norm of the initial state \eqref{initial-state-3d-firtscase} is greater than the $l_1$ norm of the final state \eqref{final-state2-3d-firtscase}. The states \eqref{initial-state-3d-firtscase} and \eqref{final-state2-3d-firtscase} form a pair of states, $\{\ket{\psi},\ket{\varphi_2}\}$, which belongs to the region $R_5$ of Fig.~\eqref{figure1}.



{\emph{The case $\tilde{\psi}_2 \leq \tilde{\varphi}_2$}}:
From the third row of Table \eqref{Table:Permutations-d234} $\{{f_{3}(k)}\}_{k=1}^{3}=\{1,3,2\}$.
Then, the terms $c_{k,n}$ and $\ket{c_{f_{n}(k)}}$ for the Kraus operator $K_3$ are given by
\begin{eqnarray}\label{D3-probablities-second-case}
c_{1,3}&=&\sqrt{p_3} \big({\varphi_1}/{\psi_1}\big), \quad \ket{c_{f_{3}(1)}}=\ket{c_1}, \nonumber \\
c_{2,3}&=&\sqrt{p_3} \big({\varphi_3}/{\psi_2}\big), \quad \ket{c_{f_{3}(2)}}=\ket{c_3}, \nonumber \\
c_{3,3}&=&\sqrt{p_3} \big({\varphi_2}/{\psi_3}\big), \quad \ket{c_{f_{3}(3)}}=\ket{c_2}.
\end{eqnarray}
This means our $(\alpha, \beta) \in [1,d]$ pair given in Eq.~\eqref{alpha-beta-permutation} is $(2,3)$.
When we perform Kraus operator $K_3$ to the initial state, we obtain $\varphi_1\ket{c_1}+\varphi_3\ket{c_3}+\varphi_2\ket{c_2}$
with probability $p_3$. 
Also the Kraus operators $F_4$, $F_5$, and $F_6$ are given by Eq.~\eqref{Kraus-d-level2}  in a way that yields $\sum_{n=1}^{3} K_{n}^{\dagger} K_{n}+\sum_{m=4}^{6} F_{m}^{\dagger} F_{m}=I$. For this case, the probabilities in Eqs.~\eqref{D3-probablities-K1}, \eqref{D3-probablities-K2}, and \eqref{D3-probablities-second-case}, are found to be
\begin{eqnarray}\label{probabilities-3d-second-case}
p_1=1-p_2-p_3, \quad
p_2=\frac{\tilde{\varphi}_1-\tilde{\psi}_1}{\tilde{\varphi}_1-\tilde{\varphi}_3}, \quad
p_3=\frac{\tilde{\varphi}_2-\tilde{\psi}_2}{\tilde{\varphi}_2-\tilde{\varphi}_3}.
\end{eqnarray}
Here again we stress that both the majorization condition \eqref{Maj-Superposition-Condition1} and CoC \eqref{Condition2} need to be satisfied (as seen, the condition for the positivity of the probabilities implies $\tilde{\psi}_1 \leq \tilde{\varphi}_1$ which is just the majorization relation and $\tilde{\psi}_2 \leq \tilde{\varphi}_2$ which is the case we are interested in), and the corresponding $\omega$ matrix is given by
\begin{eqnarray}\begin{aligned}
\omega=\left(\begin{array}{cccccc}  \varphi_1^2 & \varphi_2^2 & \varphi_3^2  \\ \varphi_3^2 & \varphi_2^2 & \varphi_1^2 \\ \varphi_1^2 & \varphi_3^2 & \varphi_2^2 \end{array}\right).
\end{aligned}\end{eqnarray}
It is easy to find examples for the regions $\{R_i\}_{i=1,\dots,5}$ where only the transformations in the region $R_1$ are deterministic.

To recap, for $d=3$, we obtain the complete solution of superposition-free transformations by discussing the problem under two cases, $\tilde{\psi}_2 \geq \tilde{\varphi}_2$ and $\tilde{\psi}_2 \leq \tilde{\varphi}_2$. One can use the solutions presented here for the desired transformations by taking notice of conditions \eqref{Maj-Superposition-Condition1} and \eqref{Condition2}.


\subsubsection{$d$-dimensional systems}

Inspired by Ref.~\cite{Torun-DetCoherence}, we follow a similar route to discuss the problem for $d$-dimensional systems.
In the following, since the problem is too complicated for $d\geq4$, we will limit ourselves to discussing how to
construct superposition-free Kraus operators.
The key point is to determine a true set of index functions $\{f_n(k)\}_{k=1}^{d}$ for $n=1,\dots,d$.
Then, constructing the superposition-free Kraus operators  $\{K_n\}_{n=1}^{d}$ given by Eq.~\eqref{Kraus-d-level1} is straightforward.

As mentioned before, for $d=3$, the relations $\tilde{\psi}_1 \leq \tilde{\varphi}_1$ and $\tilde{\psi}_3 \geq \tilde{\varphi}_3$ follow from the majorization conditions given by Eq.~\eqref{Maj-Superposition-Condition1}; but, there are two possible relations between $\tilde{\psi}_2$ and $\tilde{\varphi}_2$. Similarly, for $d$-dimensional systems, the relations $\tilde{\psi}_1 \leq \tilde{\varphi}_1$ and $\tilde{\psi}_d \geq \tilde{\varphi}_d$ follow from the majorization conditions given by Eq.~\eqref{Maj-Superposition-Condition1}. However, for the remaining coefficients we have either $\tilde{\psi}_k \geq \tilde{\varphi}_k$ or $\tilde{\psi}_k \leq \tilde{\varphi}_k$ for $k=2,3,\dots,(d-1)$, i.e., there are $2^{(d-2)}$ possible cases for $d\geq3$. By adapting the protocol presented in Ref.~\cite{Torun-DetCoherence}, all set of index functions $\{f_n(k)\}_{k=1}^{d}$ for $n=1,\dots,d$ can be easily obtained for any possible cases between the coefficient $\tilde{\psi}_k$ and  $\tilde{\varphi}_k$ for $k=2,3,\dots,(d-1)$.

In general, for coherence theory, constructing a general form of Kraus operators (i.e., constructing a general form of probabilities) is a highly nontrivial problem, and also the problem becomes exponentially difficult as the dimension gets greater \cite{Torun-DetCoherence}. The situation is clearly similar for superposition-free transformations. However, we are able to extrapolate a complete solution for some special cases of $d$-dimensional systems. Here, we give two examples. First, let us consider the case $\tilde{\psi}_k \geq \tilde{\varphi}_k$ for any $k=2,3,\dots,(d-1)$. One can obtain all $(\alpha, \beta) \in [1,d]$ pairs given in Eq.~\eqref{alpha-beta-permutation} for index functions $\{f_n(k)\}_{k=1}^{d}$. We then have a set of (permutation) pairs $\{(1,k)\}_{k=2}^d$, i.e., the order of index functions are given by
\begin{eqnarray}
& \{f_2(k)\}_{k=1}^{d} = \{2,1,3,4,5,\dots,(d-1),d\}, \quad ({\text{for}} \ \  K_2), & \nonumber \\
& \{f_3(k)\}_{k=1}^{d} = \{3,2,1,4,5,\dots,(d-1),d\}, \quad ({\text{for}} \ \  K_3), & \nonumber \\
& \{f_4(k)\}_{k=1}^{d} = \{4,2,3,1,5,\dots,(d-1),d\}, \quad ({\text{for}} \ \  K_4), & \nonumber \\
& \vdots & \nonumber \\
& \{f_d(k)\}_{k=1}^{d} = \{d,2,3,4,5,\dots,(d-1),1\}, \quad ({\text{for}} \ \  K_d), & \ \
\end{eqnarray}
where, for the superposition-free Kraus operator $K_1$, $\{f_1(k)\}_{k=1}^{d}$ $=$ $\{1,2,3\dots,d\}$.
Here, the probabilities are found to be
\begin{eqnarray}
p_1=1-\sum_{k=2}^{d} p_k, \quad
p_k=\frac{\tilde{\psi}_k-\tilde{\varphi}_k}{\tilde{\varphi}_1-\tilde{\varphi}_k},
\end{eqnarray}
where $\sum_{n=1}^{d} p_n=1$ ($p_n \geq 0$). Second, let us consider the case $\tilde{\psi}_k \leq \tilde{\varphi}_k$ for any $k=2,3,\dots,(d-1)$.
One can obtain all $(\alpha, \beta) \in [1,d]$ pairs given in Eq.~\eqref{alpha-beta-permutation} for index functions $\{f_n(k)\}_{k=1}^{d}$.
We then have $\{(k,d)\}_{k=1}^{d-1}$, i.e., the order of index functions are given by
\begin{eqnarray}
& \{f_2(k)\}_{k=1}^{d} = \{d,2,3,4,5,\dots,(d-1),1\}, \quad ({\text{for}} \ \  K_2), & \nonumber \\
& \{f_3(k)\}_{k=1}^{d} = \{1,d,3,4,5,\dots,(d-1),2\}, \quad ({\text{for}} \ \  K_3), & \nonumber \\
& \{f_4(k)\}_{k=1}^{d} = \{1,2,d,4,5,\dots,(d-1),3\}, \quad ({\text{for}} \ \  K_4), & \nonumber \\
& \vdots & \nonumber \\
& \{f_d(k)\}_{k=1}^{d} = \{1,2,3,4,5,\dots,d,(d-1)\}, \quad ({\text{for}} \ \  K_d), & \ \
\end{eqnarray}
where, for the superposition-free Kraus operator $K_1$, $\{f_1(k)\}_{k=1}^{d}$ $=$ $\{1,2,3\dots,d\}$.
Here, the probabilities are found to be
\begin{eqnarray}
p_1=1-\sum_{k=2}^{d} p_k, \quad
p_k=\frac{\tilde{\varphi}_{k-1}-\tilde{\psi}_{k-1}}{\tilde{\varphi}_{k-1}-\tilde{\varphi}_d},
\end{eqnarray}
where $\sum_{n=1}^{d} p_n=1$ ($p_n \geq 0$). These are just two examples of some of the generalizable cases. As a result, a transformation can be achieved for any given initial and final states [of course, conditions given by Eqs.~\eqref{Maj-Superposition-Condition1} and \eqref{Condition2} must be satisfied] by adapting the protocol presented in Ref.~\cite{Torun-DetCoherence} to superposition.

As we mentioned before, the conditions given by Eqs~\eqref{Maj-Superposition-Condition1} and \eqref{Condition2} presented above for a deterministic transformation also hold when the scalar products of the basis states $\braket{c_i}{c_j}=\mu_{ij}$ for $i\neq j$ are different.
Just a small change in Eq.~\eqref{Tilde-psi-varphi} is sufficient:
\begin{eqnarray}\label{Tilde-psi-varphi-general}
\tilde{\psi}_i \coloneqq \psi_i(\psi_i+ \mathop{\sum_{j=1}^{d}}_{(j\neq i)}\mu_{ij}\psi_j),
\quad
\tilde{\varphi}_i \coloneqq \varphi_i(\varphi_i+ \mathop{\sum_{j=1}^{d}}_{(j\neq i)}\mu_{ij}\varphi_j), \quad
\end{eqnarray}
for $i=1,2,\dots,d$ where the coefficients $\tilde{\psi}_i$ and $\tilde{\varphi}_i$ are in an order such that
$\tilde{\psi}_l\geq\tilde{\psi}_{l+1}$ and $\tilde{\varphi}_l\geq\tilde{\varphi}_{l+1}$ for any $l\in [1,d-1]$.
Thus, everything regarding the protocol we have introduced is the same, only Eq~\eqref{Tilde-psi-varphi} is replaced by Eq.~\eqref{Tilde-psi-varphi-general}.

\section{Maximal superposition states}\label{Sec:Maximal}

In any resource theory, a vital issue is to identify the levels of resourcefulness. In this section,
we focus on the maximally resourceful state---the state with the greatest resource value, i.e., the state at
the top of the hierarchy of resourcefulness. By the definition, a $d$-dimensional superposition state is said to
have maximal superposition if it can be used to generate all other $d$-dimensional states deterministically 
using $\mathcal{FO}$.

The existence of maximally resourceful states is well defined for the resource theory of coherence \cite{Baumgratz-Coherence},
a maximally coherent state is given by $\ket{\Psi_d}\coloneqq(1/\sqrt{d})\sum_{i=1}^{d}\ket{i}$. In analogy to the theory of coherence,
one may try to formulate maximal superposition states; but it is not trivial in general. It has been shown that such golden units
exist only for qubits in superposition theory~\cite{Plenio-RTofS}. However, thorough seeking can give us more interesting results.

In superposition theory, nonorthogonality of the basis states determines every aspect of the theory including the existence of the maximally resourceful states. Besides, it is necessary to investigate the maximal state in accordance with whether the scalar product is positive or negative. It turns out that the maximal state is a symmetric superposition of basis states for a negative scalar product.
To gain more insight into the maximal superposition states, it would be a more correct step to delve
into the resourcefulness for negative scalar products.
Keeping in mind that a state with maximal superposition has to maximize the $l_1$ norm of superposition,
we study the maximally resourceful states for negative values of scalar products. The following sections
aim to explore these kinds of states for $d\geq2$ with various examples.



\subsection{Qubit Systems}\label{Subsec:Max-qubit}

Here we present maximally resourceful superposition state(s) for two-dimensional systems. As mentioned before, one first needs to observe whether the scalar product of basis states is positive or negative.
In this sense, the state
\begin{eqnarray}\label{Max-SupState-mu-Positive}
\ket{\Psi_{-}} = \frac{1}{\sqrt{2\big(1-\mu\big)}}\Big(\ket{c_1}-\ket{c_2}\Big),
\end{eqnarray}
is maximal for $0 \leq \mu < 1$ \cite{Plenio-RTofS} and the state
\begin{eqnarray}\label{Max-SupState-mu-Negative}
\ket{\Psi_{+}} = \frac{1}{\sqrt{2\big(1+\mu\big)}}\Big(\ket{c_1}+\ket{c_2}\Big),
\end{eqnarray}
is maximal for $-1 < \mu \leq 0$, where $\mu=\braket{c_1}{c_2}$ and  $\tilde{\psi}_1=\tilde{\psi}_2=1/2$ both for $\ket{\Psi_{-}}$ and $\ket{\Psi_{+}}$.
The state given in Eq.~\eqref{Max-SupState-mu-Positive} (where $\lambda < 0$) and the state given in Eq.~\eqref{Max-SupState-mu-Negative} (where $\lambda > 0$) can be transformed to any other state $\ket{\varphi}=\varphi_1\ket{c_1}+\varphi_2\ket{c_2}$ ($|\varphi_1|\geq|\varphi_2|$)
when $\mu \in [0,1)$ and $\mu \in (-1,0]$, respectively, by using the superposition-free Kraus operators
presented above---Eqs.~\eqref{K1-qubit} and \eqref{K2-qubit}---for two-dimensional systems.

Once again, the role of the scalar product is central in considering the maximally resourceful states.
For instance, the state given in Eq.~\eqref{Max-SupState-mu-Positive} cannot be transformed
into another state (with unit probability) when $\mu$ is negative. This clearly shows that scalar product has a major impact on
superposition-free transformations. As a result, we have two sets of superposition states:
$\{\ket{\Psi_{-}}, \phi_1\ket{c_1}+\phi_2\ket{c_2}\}$ 
where $\mu \in [0,1)$ and $\{\ket{\Psi_{+}}, \chi_1\ket{c_1}+\chi_2\ket{c_2}\}$ 
where $\mu \in (-1,0]$. The state $\ket{\Psi_{-}}$ is the maximal one for the former and the state
$\ket{\Psi_{+}}$ is the maximal one for the latter.






\subsection{$d$-dimensional Systems}\label{Subsec:Max-d}

At first sight, the results obtained for qubit systems give an idea for higher dimensions; however, the problem in dimensions greater than two is more complicated. The results show us that the maximal superposition states are
\begin{eqnarray}\label{Max-SupState-ddim}
\ket{\Psi_{+}} \coloneqq \frac{1}{\sqrt{d\big(1+(d-1)\mu\big)}}\sum_{i=1}^{d}\ket{c_i},
\end{eqnarray}
where the scalar product can be $1/(1-d)<\mu\leq 0$ for $d\geq 3$ and $\mu=\braket{c_i}{c_j}$ for $i,j=1, \dots, d$ ($i\neq j$).
For the state $\ket{\Psi_+}$ we have $\tilde{\psi}_i=1/d$.
The state given in Eq.~\eqref{Max-SupState-ddim} may be transformed to another state $\ket{\varphi}=\sum_{i=1}^{d}\varphi_i\ket{c_i}$, where the coefficients $\varphi_i$ are real.
Our findings regarding the state given by Eq.~\eqref{Max-SupState-ddim} is as follows: The state given in Eq.~\eqref{Max-SupState-ddim} can be treated as ``maximally resourceful'' for target states 
$\ket{\varphi}=\sum_{i=1}^{d}\varphi_i\ket{c_i}$ where $\varphi_i \geq 0$.
Therefore, we have a set of states
\begin{eqnarray}\label{Max-superposition-set}
\Big\{\ket{\Psi_{+}}, \sum_{i=1}^{d}\varphi_i\ket{c_i}\Big\}, \quad \mu \in (\frac{1}{1-d},0],
\end{eqnarray}
where $\varphi_i \geq 0$ (a symmetric superposition of basis states).
Then the state $\ket{\Psi_{+}}$, given by Eq.~\eqref{Max-SupState-ddim}, is the ``maximally resourceful'' state of this set given by Eq.~\eqref{Max-superposition-set}. Such an approach would be reasonable for the investigation of maximally resourceful states in the resource theory of superposition.

We note that for a given target state $\ket{\varphi}=\sum_{i=1}^{d}\varphi_i\ket{c_i}$ with $\varphi_i \geq 0$ and
$1/(1-d)<\mu=\braket{c_i}{c_j}\leq 0$ it is always possible to make
$\tilde{\varphi}_k \geq \tilde{\varphi}_{k+1}$ for $k=1,2,\dots, d-1$ with
superposition-free flip operations, where the coefficients $\varphi_i \geq 0$ are not necessarily to be in decreasing or increasing order. In other words, first of all, we check whether it is provided
$\tilde{\varphi}_k \geq \tilde{\varphi}_{k+1}$ for a given target state $\sum_{i=1}^{d}\varphi_i\ket{c_i}$, and then, we apply our protocol.

To elucidate the above discussion, we now give explicit examples for $d=3,4$; two examples for $d=3$ and one for $d=4$.
First, let us consider the case $d=3$ where the maximally resourceful state is given by
\begin{eqnarray}\label{max-super-d3}
\ket{\Psi_{+}} = \frac{1}{\sqrt{3\big(1+2\mu\big)}}\sum_{i=1}^{3}\ket{c_i}, \quad \big(\mu \in (-\frac{1}{2},0]\big),
\end{eqnarray}
where  $\tilde{\psi}_1=\tilde{\psi}_2=\tilde{\psi}_3=1/3$. For $d=3$, one can find examples where only either the solutions of the case $\tilde{\psi}_2=1/3 \geq \tilde{\varphi}_2$ or $\tilde{\psi}_2=1/3 \leq \tilde{\varphi}_2$ can be used. Also, there are examples where these two cases work for a specific range of scalar product separately; that is, $-{1}/{2}<\mu\leq \alpha$ works for one of these cases and  $\alpha<\mu\leq0$ works for the other case. Consider, for instance, the given target state
\begin{eqnarray}\label{Max-d3-example1}
\ket{\varphi}=\frac{1}{\sqrt{45+76\mu}}
\Big(5\ket{c_1}+4\ket{c_2}+2\ket{c_3}\Big),
\end{eqnarray}
where $-1/2<\mu\leq 0$.
The basic outline of the path to be followed is simple.
First, find $\tilde{\psi}_2$ and $\tilde{\varphi}_2$ for the given (initial and final) states; $\tilde{\psi}_2=1/3$ and $\tilde{\varphi}_2=(16+28\mu)/(45+76\mu)$. Second, find the range of $\mu$ for each case $\tilde{\psi}_2 \geq \tilde{\varphi}_2$ and $\tilde{\psi}_2 \leq \tilde{\varphi}_2$; $-1/2 < \mu \leq -0.375$ works for the former and $-0.375\leq \mu \leq0$ works for the latter. Here, it is easy to check that
$\tilde{\varphi}_1 \geq \tilde{\varphi}_{2} \geq \tilde{\varphi}_{3}$ for both cases. Next, check the  majorization condition given by Eq.~\eqref{Maj-Superposition-Condition1} and CoC given by Eq.~\eqref{Condition2} to be satisfied. Then, the transformation $\ket{\Psi_{+}}\overset{\mathcal{FO}}{\longrightarrow}\ket{\varphi}$ can be achieved for an arbitrary $\mu \in (-\frac{1}{2},0]$ by using the solutions of three-dimensional systems. For instance, let us take $\mu=-9/19$ in Eq.~\eqref{Max-d3-example1}, and then the target state is
\begin{eqnarray}
\ket{\varphi}=\frac{5}{3}\ket{c_1}+\frac{4}{3}\ket{c_2}
+\frac{2}{3}\ket{c_3}.
\end{eqnarray}
From the solutions of the case $\tilde{\psi}_2 \geq \tilde{\varphi}_2$ ($\tilde{\psi}_2=1/3$ and $\tilde{\varphi}_2=52/171$) of three-dimensional systems, superposition-free Kraus operators are given by
\begin{eqnarray}\label{K1-3d-1st}
	K_1=\frac{\sqrt{p_1}}{\sqrt{57}}
	\Big(5\frac{\ket{c_1}\bra{c_1^{\perp}}}{\zeta_1}
	+4\frac{\ket{c_2}\bra{c_2^{\perp}}}{\zeta_2}
	+2\frac{\ket{c_3}\bra{c_3^{\perp}}}{\zeta_3}\Big), \quad
\end{eqnarray}
\begin{eqnarray}\label{K2-3d-1st}
	K_2=\frac{\sqrt{p_2}}{\sqrt{57}}
	\Big(2\frac{\ket{c_3}\bra{c_1^{\perp}}}{\zeta_1}
	+4\frac{\ket{c_2}\bra{c_2^{\perp}}}{\zeta_2}
	+5\frac{\ket{c_1}\bra{c_3^{\perp}}}{\zeta_3}\Big), \quad
\end{eqnarray}
\begin{eqnarray}\label{K3-3d-1st}
	K_3=\frac{\sqrt{p_3}}{\sqrt{57}}
	\Big(4\frac{\ket{c_2}\bra{c_1^{\perp}}}{\zeta_1}
	+5\frac{\ket{c_1}\bra{c_2^{\perp}}}{\zeta_2}
	+2\frac{\ket{c_3}\bra{c_3^{\perp}}}{\zeta_3}\Big), \quad
\end{eqnarray}
where $\zeta_i=\braket{c_i^{\perp}}{c_i}$ and $K_i\ket{\Psi_+}=\sqrt{p_i}\ket{\varphi}$ for $i=1,2,3$. The probabilities
for the respective outcomes are $p_1=\frac{7063}{14841}$, $p_2=\frac{143}{291}$, and $p_3=\frac{5}{153}$, where $p_1+p_2+p_3=1$ indeed. It is also easy to check that the CoC given by Eq.~\eqref{F3F4-qubit-explicit} is satisfied.

Second, consider the following example for $d=3$ where the final state is given by
\begin{eqnarray}\label{phi-flip-example}
\ket{\phi}=\sqrt{\frac{5}{593}}
\Big(12\ket{c_1}+2\ket{c_2}+\ket{c_3}\Big),
\end{eqnarray}
with $\mu=-2/5$. It is easy to check that
$\tilde{\phi}_1 \geq \tilde{\phi}_{2} \geq \tilde{\phi}_{3}$ is not satisfied. However, with the superposition-free flip operation between basis states $\ket{c_2}$ and $\ket{c_3}$, the state given in Eq.~\eqref{phi-flip-example} can be written such that
\begin{eqnarray}\label{phi-flip-example2}
	\ket{\phi}=\sqrt{\frac{5}{593}}
	\Big(12\ket{c_1}+\ket{c_2}+2\ket{c_3}\Big).
\end{eqnarray}
Now it is straightforward to check that $\tilde{\phi}_1 \geq \tilde{\phi}_{2} \geq \tilde{\phi}_{3}$ is satisfied for the state given in Eq.~\eqref{phi-flip-example2}. Since the state given in Eq.~\eqref{phi-flip-example} could be transformed into the state given in Eq.~\eqref{phi-flip-example2} by superposition-free flip operations, they are equivalent states in the context of resource theory of superposition. Then, by using the solutions of the case $\tilde{\psi}_2 \geq \tilde{\varphi}_2$ ($\tilde{\psi}_2=1/3$ and $\tilde{\phi}_2=-23/593$) of three-dimensional systems the transformation $\ket{\Psi_{+}}\overset{\mathcal{FO}}{\longrightarrow}\ket{\phi}$ can be achieved deterministically.

Third, consider the case $d=4$ where the maximally resourceful state is given by
\begin{eqnarray}\label{max-super-d4}
\ket{\Psi_{+}} = \frac{1}{\sqrt{4\big(1+3\mu\big)}}\sum_{i=1}^{4}\ket{c_i}, \quad \big(\mu \in (-\frac{1}{3},0]\big).
\end{eqnarray}
There are five possible cases for $d=4$ as seen from Table \eqref{Table:Permutations-d234} (or from Ref.~\cite{Torun-DetCoherence} for coherence). Consider, for instance, the given final state
\begin{eqnarray}
\ket{\chi}=\frac{1}{\sqrt{110+214\mu}}\Big(9\ket{c_1}+4\ket{c_2}+3\ket{c_3}+2\ket{c_4}\Big).
\end{eqnarray}
For $-1/3< \mu \leq0$ the transformation $\ket{\Psi_{+}}\overset{\mathcal{FO}}{\longrightarrow}\ket{\chi}$ can be achieved deterministically by constructing Kraus operators with the help of Table \eqref{Table:Permutations-d234}. The case $\tilde{\psi}_2 \geq \tilde{\chi}_2$ and $\tilde{\psi}_3 \geq \tilde{\chi}_3$ (corresponding to the fourth row of Table \eqref{Table:Permutations-d234}) works for
$\mu \in (-\frac{1}{3},0]$. In higher dimensions, the problem becomes even more complex; however, similar steps would be followed to achieve the desired superposition-free transformations.

In summary, a hierarchy can be defined among pure superposition states by classifying the states according to
the range of scalar product, i.e., whether $\mu$ is negative or positive. This classification leads to a set of states given by
Eq.~\eqref{Max-superposition-set} where the state given by Eq.~\eqref{Max-SupState-ddim} is the ``maximally resourceful'' state.
This way of thinking about resourcefulness allows us to partially explore the existence of maximal superposition states.

\section{Conclusion}\label{Sec:Conc}

In this work, inspired by Ref.~\cite{Torun-DetCoherence}, we have developed an explicit framework for the manipulation of
superposition states as being one of the central problems of the resource theory of superposition \cite{Plenio-RTofS}.
For this purpose, we first have provided superposition-free operators for a deterministic transformation.
Moreover, we have presented the conditions for a class of superposition state transformations. These conditions strictly depend on the scalar products of the basis states and reduce to the well-known majorization condition for quantum coherence \cite{Du-CoherenceMaj} in the limit of orthonormal basis. Along the way, we have completely solved the problem for $d=2,3$ and discussed for $d\geq 4$ how to construct superposition-free Kraus operators for the desired transformations by adapting the protocol introduced in Ref~\cite{Torun-DetCoherence}.

We further have expanded our study by examining the maximally resourceful states. We have determined the maximal
superposition states which are valid over a certain range of scalar products. Importantly, we have observed that
the state with the symmetric superposition of the basis states, where the scalar product is negative, can be
treated as maximal for a given particular set of states. We have explicitly discussed this problem
for the cases $d=2,3,4$ with various examples. More broadly, research is also needed to determine the resourceful states for
high-dimensional systems especially in the case of positive scalar products.

By providing conditions for a class of superposition-free transformations, our work paves the way for the investigation of mixed state transformations, transformations in the asymptotic limit \cite{Winter120404}, approximate transformations \cite{Renes_RT}, and catalytic transformations \cite{Aberg_CC,Kaifeng_CC}. Any efforts in this direction would be worthwhile, particularly because this leads to a hierarchy structure that characterizes the potential usefulness of resource states in information processing tasks.




\begin{acknowledgments}
We thank Onur Pusuluk and Ferruh \.{I}lhan for fruitful discussions.
G.T. is partially supported by the Bo{\u{g}}azi{\c{c}}i University Research Fund under Grant No. 20B03SUP3.

G.T. and H.T.\c{S}. contributed equally to this work.
\end{acknowledgments}

\renewcommand{\appendixname}{APPENDIX}
\appendix

\section{MAJORIZATION AND GRAM MATRIX}\label{Sec:App-MatrixApr}

It is known that a vector $x$ is majorized by another vector $y$ (equivalently $y$ majorizes $x$), written $x \prec y$, if and only if $x = Dy$ where $D$ is a doubly stochastic matrix \cite{Bhatia,Marshall}. In the resource theory coherence, the majorization condition for a deterministic transformation of coherent states under incoherent operations \cite{Du-CoherenceMaj} is given as
$ (\psi_1^2, \psi_2^2) \prec (\phi_1^2, \phi_2^2) $ (in dimension two) or in terms of doubly stochastic matrix
\begin{eqnarray}
    \begin{pmatrix}
        \psi_1^2 \\
        \psi_2^2  \\
     \end{pmatrix}
     =
     D
     \begin{pmatrix}
        \phi_1^2 \\
        \phi_2^2  \\
     \end{pmatrix}.
\end{eqnarray}
Additionally, in Ref.~\cite{Plenio-RTofS} it was shown that the superposition-free operators can be written as
$K_n = V \tilde{K}_n V^{-1} $ where $\tilde{K}_n$ is an incoherent operator and $V$ is a basis transformation matrix. The condition to be a trace preserving operation is given by $\sum_{n} (V^\dagger)^{-1} \tilde{K}_n^{\dagger} V^{\dagger} V \tilde{K}_n V^{-1}=I$. 
By multiplying this condition with $ V^{\dagger} $ and $V$ from left and right, respectively, it becomes $\sum_{n} \tilde{K}_n^{\dagger} G  \tilde{K}_n = G$. Since off-diagonal terms of the Gram matrix $G_{ij} = \mu_{ij}$, in the limit of orthonormal basis $G \rightarrow I $, this suggests a continuity in majorization condition due to the scalar product $\mu$ as we obtained in Eq.~\eqref{Maj-Superposition-Condition1}.
Then, in the resource theory of superposition, the majorization condition given by Eq.~\eqref{Maj-Superposition-Condition1} can be written as $ (\psi_1^2 + \mu \psi_1 \psi_2, \psi_2^2 + \mu \psi_1 \psi_2) \prec (\phi_1^2 + \mu \phi_1 \phi_2, \phi_2^2 + \mu \phi_1 \phi_2)$ for $d=2$ or in terms of doubly stochastic matrix
\begin{eqnarray}
    \begin{pmatrix}
        \psi_1^2 + \mu \psi_1 \psi_2 \\
        \psi_2^2 + \mu \psi_1 \psi_2 \\
     \end{pmatrix}
     =
     D
     \begin{pmatrix}
        \phi_1^2 + \mu \varphi_1 \varphi_2 \\
        \phi_2^2 + \mu \varphi_1 \varphi_2 \\
     \end{pmatrix}.
\end{eqnarray}
Furthermore, the above equation can be decomposed into
\begin{eqnarray}\label{Major_matrix}
    \begin{pmatrix}
        \psi_1 & 0  \\
        0      & \psi_2  \\
     \end{pmatrix}
     G
     \begin{pmatrix}
        \psi_1  \\
        \psi_2  \\
     \end{pmatrix}
     =
     D
     \begin{pmatrix}
        \phi_1 & 0  \\
        0      & \phi_2  \\
     \end{pmatrix}
     G
     \begin{pmatrix}
        \phi_1  \\
        \phi_2  \\
     \end{pmatrix},
\end{eqnarray}
where $G$ is the Gram matrix given in Eq.~\eqref{gram-matrix}. It is now obvious that the majorization condition given by Eq.~\eqref{Maj-Superposition-Condition1} or Eq.~\eqref{Major_matrix} is the generalized version of the majorization obtained in the coherence theory, i.e., in the limit of orthonormal basis, the Gram matrix $G \rightarrow I$, the above expression reduces to $ (\psi_1^2, \psi_2^2) \prec (\phi_1^2, \phi_2^2)$. Even though we have shown it for a qubit system explicitly, it is straightforward to show that this relation holds for an arbitrary dimension of Hilbert space.


\section{DERIVATION OF THE CONDITIONS FOR DETERMINISTIC TRANSFORMATIONS}\label{Sec:App-ConditionsProof}

In this Appendix, we show how to obtain the majorization condition for superposition given by
Eq.~\eqref{Maj-Superposition-Condition1} and CoC given by Eq.~\eqref{Condition2}. All the details
are hidden in the completeness relation given by Eq.~\eqref{Completeness-KnFm}, and can be derived
by a careful calculation.

As stated in Ref.~\cite{Plenio-RTofS}, assume one has an (incomplete) set of Kraus operators $\{K_n\}$ such that
$\sum_{n} K_n^{\dagger}K_n \leq I$. Then, it was proven \cite{Plenio-RTofS} that there always exist superposition-free
Kraus operators $\{F_m\}$ with $\sum_{n} K_n^{\dagger}K_n + \sum_{m} F_m^{\dagger}F_m = I$. Here, the identity operator
can be represented in the following way:
\begin{eqnarray}
I=\sum_{i,j} \frac{\ket{c_i^{\perp}}\bra{c_j^{\perp}}}{\braket{c_i}{c_i^{\perp}}\braket{c_j^{\perp}}{c_j}}\braket{c_i}{c_j}.
\end{eqnarray}
Additionally, it is not necessary to obtain the Kraus operators $\{F_m\}$ explicitly; however, the completeness relation
provides constraint(s) on superposition-free Kraus operators $\{F_m\}$, i.e., on the terms of Kraus operators given by
Eq.~\eqref{Kraus-d-level2}. The CoC given by Eq.~\eqref{Condition2} is obtained as a result of these constraints.

Using the same notation used in Sec.~\eqref{Subsec:Deterministic} for the Kraus operators
given by Eq.~\eqref{Kraus-d-level2}, we first define
\begin{eqnarray}\label{App:B-cj-positivity}
\sum_{m=d+1}^{2d} c_{j,m}^2 \coloneqq X_j,
\end{eqnarray}
for $j=2,3,\dots,d$, and
\begin{eqnarray}\label{App:B-cjl}
\sum_{m=d+1}^{2d} c_{j,m} c_{l,m} \coloneqq Y_{jl},
\end{eqnarray}
for $j=2,3,\dots,d$, $l=3,4,\dots,d$, and $j<l$. The completeness relation given by Eq.~\eqref{Completeness-KnFm}
gives us $d(d+1)/2$ equations. We divide these equations into three separate groups by combining them with
Eqs.~\eqref{App:B-cj-positivity} and \eqref{App:B-cjl}.

The first group consists of $(d-2)(d-1)/2$ equations which give us the terms $Y_{jl}$ defined by Eq~\eqref{App:B-cjl}
in terms of $\{\psi_i\}_{i=1}^d$, $\{\varphi_i\}_{i=1}^d$, $\mu$, and $\{p_n\}_{n=1}^d$. In addition, the terms $Y_{jl}$
can be either positive or negative. Therefore, the terms $Y_{jl}$ defined by Eq.~\eqref{App:B-cjl} do not indicate any
condition (or constraint).

The second group consists of $(d-1)$ equations which give us the terms $\{X_j\}_{n=2}^d$ defined by Eq.~\eqref{App:B-cj-positivity}.
Equations in this group also contain the terms $Y_{jl}$ defined by Eq.~\eqref{App:B-cjl}. By combining the equations in the first group
and second group together in a way that we obtain the terms $\{X_j\}_{n=2}^d$ in terms of $\{\psi_i\}_{i=1}^d$, $\{\varphi_i\}_{i=1}^d$,
$\mu$, and $\{p_n\}_{n=1}^d$, it is obvious that the terms $\{X_j\}_{n=2}^d$ defined by Eq.~\eqref{App:B-cj-positivity} must be non-negative.
Therefore, the positivity of these terms implies the CoC.

The third group consists of $d$ equations where the unknowns are probabilities $\{p_n\}_{n=1}^d$. Equations in this group also
contain the terms $Y_{jl}$ and $X_{j}$. These $d$ equations can be reduced into the form given by Eq.~\eqref{probabilities-d-dim} by using the definitions given by Eqs.~\eqref{Tilde-psi-varphi}, \eqref{App:B-cj-positivity}, and \eqref{App:B-cjl}. By solving these, one obtains the probabilities in terms of just $\{\tilde{\psi}_i\}_{i=1}^d$ and $\{\tilde{\varphi}_i\}_{i=1}^d$. We show that Eq.~\eqref{probabilities-d-dim}
can be written in a way that contains a doubly stochastic matrix which provides us the majorization condition for superposition.

In what follows, we obtain the majorization condition given by Eq.~\eqref{Maj-Superposition-Condition1} and the CoC given by Eq.~\eqref{Condition2} for $d=2,3$ and give the procedure to be applied for arbitrary dimension. Now, first consider the simplest case, qubit systems. The index functions for the Kraus operators $K_1$, $K_2$, $F_3$, and $F_4$ are given such that $\{f_1(1),f_1(2)\} = \{1,2\}$, $\{f_2(1),f_2(2)\} = \{2,1\}$, $\{f_3(1),f_3(2)\} = \{1,1\}$, and $\{f_4(1),f_4(2)\} = \{2,2\}$, respectively. After constructing the Kraus operators given by Eqs.~\eqref{Kraus-d-level1} and \eqref{Kraus-d-level2}, the completeness relation given by Eq.~\eqref{Completeness-KnFm} gives us the following three equations:
\begin{eqnarray}\label{App-d2-eq1}
c_{1,1}^2 + c_{1,2}^2 + c_{1,3}^2 + c_{1,4}^2 = 1,
\end{eqnarray}
\begin{eqnarray}\label{App-d2-eq2}
c_{2,1}^2 + c_{2,2}^2 + c_{2,3}^2 + c_{2,4}^2 = 1,
\end{eqnarray}
\begin{eqnarray}\label{App-d2-eq3}
\big(c_{1,1}c_{2,1} + c_{1,2}c_{2,2}\big)\mu + c_{1,3}c_{2,3} + c_{1,4}c_{2,4} = \mu.
\end{eqnarray}
As we mentioned above, we divide these equations into three groups. Here, there is no equation in the first group. In the second group we have only one equation, Eq.~\eqref{App-d2-eq3}, and in the third group we have two equations, Eqs.~\eqref{App-d2-eq1} and \eqref{App-d2-eq2}.
Also, from $K_n\ket{\psi} = \sqrt{p_n}\ket{\varphi}$ we have
\begin{eqnarray}\label{App:B-Kraus-K1K2-d2-condition}
c_{1,1} &=& \sqrt{p_1} (\varphi_1 / \psi_1 ), \quad c_{2,1} = \sqrt{p_1} (\varphi_2 / \psi_2), \nonumber \\
c_{1,2} &=& \sqrt{p_2} (\varphi_2 / \psi_1 ), \quad c_{2,2} = \sqrt{p_2} (\varphi_1 / \psi_2),
\end{eqnarray}
and from $F_m\ket{\psi} = 0 $ we have
\begin{eqnarray}\label{App:B-Kraus-F3F4-d2-condition}
c_{1,3} = -c_{2,3} (\psi_2/\psi_1), \quad  c_{1,4} = -c_{2,4} (\psi_2/\psi_1).
\end{eqnarray}
From Eq.~\eqref{App-d2-eq3} we obtain the CoC for qubit systems. Obviously, Eq.~\eqref{App:B-cj-positivity} for $d=2$ gives us
\begin{eqnarray}\label{App:B-cj-positivity-d2}
X_2=c_{2,3}^2+c_{2,4}^2.
\end{eqnarray}
Combining Eqs.~\eqref{App:B-Kraus-K1K2-d2-condition}, \eqref{App:B-Kraus-F3F4-d2-condition}, and \eqref{App:B-cj-positivity-d2} with Eq.~\eqref{App-d2-eq3}, we obtain
\begin{eqnarray}\label{App:B-X2-dim2}
X_2=\frac{1}{\psi_2^2}\Big(\varphi_1\varphi_2\mu - \psi_1\psi_2\mu\Big).
\end{eqnarray}
Here, the term $X_2$ is non-negative; then the right-hand side of Eq.~\eqref{App:B-X2-dim2} must be non-negative, yielding
\begin{eqnarray}\label{App:B-CoC-firstform}
\mu(\varphi_1\varphi_2-\psi_1\psi_2)\geq 0.
\end{eqnarray}
Moreover, it is useful to obtain a more compact form of Eq.~\eqref{App:B-CoC-firstform}, which serves as a guide to finding out the CoC given by Eq.~\eqref{Condition2} for higher-dimensional systems. We return to this point after we obtain the probabilities $p_1$ and $p_2$.

As mentioned before, the explicit construction of $F_3$ and $F_4$ is not necessary \cite{Plenio-RTofS}. Furthermore, while $X_2$ given in Eq.~\eqref{App:B-X2-dim2} is equal to $c_{2,3}^2+c_{2,4}^2$, different choices of $c_{2,3}$ and $c_{2,4}$ give us different sets of $\{F_3, F_4\}$, provided that Eq.~\eqref{App:B-X2-dim2} is satisfied. For instance, if we choose $c_{2,4}=0$ (or $c_{2,3}=0$) then we have only three superposition-free Kraus operators, $\{K_1,K_2, F_3\}$ (or $\{K_1,K_2, F_4\}$). On the other hand, for $c_{2,3}\neq 0$ and $c_{2,4}\neq 0$, we need four superposition-free Kraus operators to make the entire operation trace preserving.

We now proceed with equations in the third group.
Combining Eqs.~\eqref{App:B-Kraus-K1K2-d2-condition}, \eqref{App:B-Kraus-F3F4-d2-condition}, \eqref{App:B-cj-positivity-d2}, \eqref{App:B-X2-dim2}, and Eq.~\eqref{Tilde-psi-varphi} with equations in the third group, i.e.,  Eqs.~\eqref{App-d2-eq1} and \eqref{App-d2-eq2}, we get
\begin{eqnarray}\label{APP:B-probabilities-1st-eq}
p_1\tilde{\varphi}_1 + p_2\tilde{\varphi}_2 = \tilde{\psi}_1,
\end{eqnarray}
\begin{eqnarray}\label{APP:B-probabilities-2nd-eq}
p_1\tilde{\varphi}_2 + p_2\tilde{\varphi}_1 = \tilde{\psi}_2.
\end{eqnarray}
The above equations can be written in the compact form
\begin{eqnarray}\label{App:B-TransformationMatrix}
    \begin{pmatrix}
        p_1 & p_2       \\
        p_2 & p_1
    \end{pmatrix}
    \begin{pmatrix}
        \tilde{\varphi}_1 \\
        \tilde{\varphi}_2
    \end{pmatrix}
    =
    \begin{pmatrix}
        \tilde{\psi}_1 \\
        \tilde{\psi}_2
    \end{pmatrix},
\end{eqnarray}
where $p_1, p_2 \geq 0$ and $p_1 + p_2 = 1$. The transformation matrix given in Eq.~\eqref{App:B-TransformationMatrix} is a doubly stochastic matrix: $D (\tilde{\varphi}_1, \tilde{\varphi}_2)^T =(\tilde{\psi}_1, \tilde{\psi}_2)^T$. Hence, the vector $(\tilde{\psi}_1, \tilde{\psi}_2)^T$ is majorized by the vector $(\tilde{\varphi}_1, \tilde{\varphi}_2)^T$ written $(\tilde{\psi}_1, \tilde{\psi}_2)^T  \prec (\tilde{\varphi}_1, \tilde{\varphi}_2)^T$, where $\tilde{\psi}_1 \geq \tilde{\psi}_2$  and $\tilde{\varphi}_1 \geq \tilde{\varphi}_2$.

Let us now return to Eq.~\eqref{App:B-CoC-firstform}.
Combining Eq.~\eqref{App:B-CoC-firstform} with Eq.~\eqref{Tilde-psi-varphi}, the inequality given in Eq.~\eqref{App:B-CoC-firstform} can be written as
$[({\psi}_2^2-{\varphi}_2^2)-(\tilde{\psi}_2-\tilde{\varphi}_2)]\geq 0$. Also, by solving Eqs.~\eqref{APP:B-probabilities-1st-eq} and \eqref{APP:B-probabilities-2nd-eq}, the probabilities are found to be
\begin{eqnarray}\label{App:B-prob-d2}
	p_1=\frac{\tilde{\varphi}_1-\tilde{\psi}_2}
	{\tilde{\varphi}_1-\tilde{\varphi}_2}, \quad p_2=\frac{\tilde{\psi}_2-\tilde{\varphi}_2}
	{\tilde{\varphi}_1-\tilde{\varphi}_2}.
\end{eqnarray}
Then, combining $[({\psi}_2^2-{\varphi}_2^2)-(\tilde{\psi}_2-\tilde{\varphi}_2)]\geq 0$ with Eq.~\eqref{App:B-prob-d2}, the inequality given in Eq.~\eqref{App:B-CoC-firstform} can be written such that
\begin{eqnarray}\label{App:B-CoC-Qubit}
	p_1\varphi_2^2+p_2\varphi_1^2 \leq \psi_2^2,
\end{eqnarray}
which gives us the CoC (for qubit systems) defined by Eq.~\eqref{Condition2}.
Furthermore, in the limit of orthonormal basis, equality holds in Eq.~\eqref{App:B-CoC-Qubit}, and also Eqs.~\eqref{APP:B-probabilities-2nd-eq} and \eqref{App:B-CoC-Qubit}  become same, i.e., majorization is the only condition, which is necessary and sufficient, for the deterministic coherence transformations.

Second, consider the case $\tilde{\psi}_2 \geq \tilde{\varphi}_2$ of $d=3$.
After constructing the Kraus operators (by using the given index functions given in Table \eqref{Table:Permutations-d234}),
the completeness relation given by Eq.~\eqref{Completeness-KnFm} gives us the following six equations:
\begin{eqnarray}\label{App-d3-eq123}
\sum_{i=1}^{6} c_{j,i}^2 = 1,
\end{eqnarray}
for $j=1,2,3$,
\begin{eqnarray}\label{App-d3-eq4}
\mu \sum_{i=1}^{3} c_{1,i} c_{2,i} + \sum_{i=4}^{6} c_{1,i} c_{2,i} = \mu,
\end{eqnarray}
\begin{eqnarray}\label{App-d3-eq5}
\mu \sum_{i=1}^{3} c_{1,i} c_{3,i} + \sum_{i=4}^{6} c_{1,i} c_{3,i} = \mu,
\end{eqnarray}
\begin{eqnarray}\label{App-d3-eq6}
\mu \sum_{i=1}^{3} c_{2,i} c_{3,i} + \sum_{i=4}^{6} c_{2,i} c_{3,i} = \mu.
\end{eqnarray}
Here, we again divide these equations into three groups. There is one equation in the first group, Eq.~\eqref{App-d3-eq6};
in the second group we have two equations, Eqs.~\eqref{App-d3-eq4} and \eqref{App-d3-eq5}; and
in the third group we have three equations given in Eq.~\eqref{App-d3-eq123}.
Also, from $F_m\ket{\psi} = 0 $ we have
\begin{eqnarray}\label{App:B-Kraus-F4F5F6-d3-condition}
c_{1,i} = -\big(c_{2,i}\psi_2 + c_{3,i}\psi_3\big)/\psi_1,
\end{eqnarray}
for $i=4,5,6$. We start with the equation in the first group.
Equation \eqref{App:B-cjl} for $d=3$ gives us
\begin{eqnarray}\label{App:B-cjl-d3}
Y_{23}=c_{2,4}c_{3,4}+c_{2,5}c_{3,5}+c_{2,6}c_{3,6}.
\end{eqnarray}
Obviously, combining Eqs.~\eqref{D3-probablities-K1}, \eqref{D3-probablities-K2}, \eqref{D3-probablities-first-case}, and \eqref{App:B-cjl-d3} with Eq.~\eqref{App-d3-eq6} we obtain
\begin{widetext}
\begin{eqnarray}\label{App:B-Y23-dim3-1st}
Y_{23}=\Big(\frac{\mu}{\psi_2\psi_3}\Big)\big(\psi_2\psi_3-p_1\varphi_2\varphi_3-p_2\varphi_1\varphi_2-p_3\varphi_1\varphi_3\big).
\end{eqnarray}
The term $Y_{23}$ given above can be either positive or negative, and therefore, provides no condition.
We now proceed with equations in the second group, i.e.,  Eqs.~\eqref{App-d3-eq4} and \eqref{App-d3-eq5}. Equation \eqref{App:B-cj-positivity} for $d=3$ gives us
\begin{eqnarray}\label{App:B-cj-positivity-d3}
X_2=c_{2,4}^2+c_{2,5}^2+c_{2,6}^2, \quad
X_3=c_{3,4}^2+c_{3,5}^2+c_{3,6}^2.
\end{eqnarray}
By suitably combining Eqs.~\eqref{D3-probablities-K1}, \eqref{D3-probablities-K2}, \eqref{D3-probablities-first-case}, \eqref{App:B-Kraus-F4F5F6-d3-condition}, and \eqref{App:B-cj-positivity-d3} with Eqs.~\eqref{App-d3-eq4} and \eqref{App-d3-eq5}, we obtain
\begin{eqnarray}\label{App:B-X2-dim3-1st}
X_2=\big(\frac{\mu}{\psi_2^2}\big)\Big(p_1[\varphi_1\varphi_2+\varphi_2\varphi_3]+ p_2[\varphi_1\varphi_2+\varphi_2\varphi_3]
+p_3[\varphi_1\varphi_2+\varphi_1\varphi_3] - [\psi_1\psi_2+\psi_2\psi_3]\Big),
\end{eqnarray}
\begin{eqnarray}\label{App:B-X3-dim3-1st}
X_3=\big(\frac{\mu}{\psi_3^2}\big)\Big(p_1[\varphi_1\varphi_3+\varphi_2\varphi_3]+ p_2[\varphi_1\varphi_2+\varphi_1\varphi_3]
+p_3[\varphi_1\varphi_3+\varphi_2\varphi_3] - [\psi_1\psi_3+\psi_2\psi_3]\Big).
\end{eqnarray}
\end{widetext}
Here, the terms $X_2$ and $X_3$ are non-negative, then the right hand side of Eqs.~\eqref{App:B-X2-dim3-1st} and \eqref{App:B-X3-dim3-1st} must be non-negative. As we discussed for qubit systems, for the compact form of CoC, which is the result of the positivity of Eq.~\eqref{App:B-X2-dim3-1st} and Eq.~\eqref{App:B-X3-dim3-1st}, we need to obtain the probabilities $p_1$, $p_2$, and $p_3$. We return this point after we obtain these probabilities.

Now, we look at equations in the third group.
Combining Eqs.~\eqref{D3-probablities-K1}, \eqref{D3-probablities-K2}, \eqref{D3-probablities-first-case}, \eqref{App:B-Kraus-F4F5F6-d3-condition}, \eqref{App:B-Y23-dim3-1st}, \eqref{App:B-X2-dim3-1st}, and \eqref{App:B-X3-dim3-1st} with three equations
given in Eq.~\eqref{App-d3-eq123}, we get
\begin{eqnarray}\label{APP:B-3dim-1st-probabilities-1st-eq}
p_1\tilde{\varphi}_1 + p_2\tilde{\varphi}_3 + p_3\tilde{\varphi}_2 = \tilde{\psi}_1,
\end{eqnarray}
\begin{eqnarray}\label{APP:B-3dim-1st-probabilities-2nd-eq}
p_1\tilde{\varphi}_2 + p_2\tilde{\varphi}_2 + p_3\tilde{\varphi}_1 = \tilde{\psi}_2,
\end{eqnarray}
\begin{eqnarray}\label{APP:B-3dim-1st-probabilities-3rd-eq}
p_1\tilde{\varphi}_3 + p_2\tilde{\varphi}_1 + p_3\tilde{\varphi}_3 = \tilde{\psi}_3.
\end{eqnarray}
The above equations can be written in the compact form
\begin{eqnarray}\label{App:B-TransformationMatrix-dim3-1st}
    \begin{pmatrix}
        p_1      & p_3         & p_2  \\
        p_3      & p_1 + p_2   & 0  \\
        p_2      & 0           & p_1 + p_3
    \end{pmatrix}
    \begin{pmatrix}
        \tilde{\varphi}_1 \\
        \tilde{\varphi}_2 \\
        \tilde{\varphi}_3
    \end{pmatrix}
    =
    \begin{pmatrix}
        \tilde{\psi}_1 \\
        \tilde{\psi}_2 \\
        \tilde{\psi}_3
    \end{pmatrix}.
\end{eqnarray}
where $p_i \geq 0$ and $\sum_{i=1}^{3} p_i = 1$. The transformation matrix given in Eq.~\eqref{App:B-TransformationMatrix-dim3-1st} is a doubly stochastic matrix: $D (\tilde{\varphi}_1, \tilde{\varphi}_2, \tilde{\varphi}_3)^T =(\tilde{\psi}_1, \tilde{\psi}_2, \tilde{\psi}_3)^T$. Hence, the vector $(\tilde{\psi}_1, \tilde{\psi}_2, \tilde{\psi}_3)^T$ is majorized by the vector $(\tilde{\varphi}_1, \tilde{\varphi}_2, \tilde{\varphi}_3)^T$ written $(\tilde{\psi}_1, \tilde{\psi}_2, \tilde{\psi}_3)^T \prec (\tilde{\varphi}_1, \tilde{\varphi}_2, \tilde{\varphi}_3)^T$, where $\tilde{\psi}_1 \geq \tilde{\psi}_2 \geq \tilde{\psi}_3$ and $\tilde{\varphi}_1 \geq \tilde{\varphi}_2 \geq \tilde{\varphi}_3$ (and also the case is $\tilde{\psi}_2 \geq \tilde{\varphi}_2$).

Let us return to Eqs.~\eqref{App:B-X2-dim3-1st} and \eqref{App:B-X3-dim3-1st}. As we mentioned above, positivity of these equations gives us the CoC given by Eq.~\eqref{Condition2}. By solving Eqs.~\eqref{APP:B-3dim-1st-probabilities-1st-eq}, \eqref{APP:B-3dim-1st-probabilities-2nd-eq}, and \eqref{APP:B-3dim-1st-probabilities-3rd-eq}, the probabilities are found to be
\begin{eqnarray}\label{App:B-probabilities-3d-first-case}
p_1=1-p_2-p_3, \quad
p_2=\frac{\tilde{\psi}_3-\tilde{\varphi}_3}{\tilde{\varphi}_1-\tilde{\varphi}_3}, \quad
p_3=\frac{\tilde{\psi}_2-\tilde{\varphi}_2}
{\tilde{\varphi}_1-\tilde{\varphi}_2}. \qquad
\end{eqnarray}
Then, combining the positivity of Eq.~\eqref{App:B-X2-dim3-1st} and Eq.~\eqref{App:B-X3-dim3-1st} with Eq.~\eqref{App:B-probabilities-3d-first-case}, we get
\begin{eqnarray}\label{App:B-CoC1-1st-3dim}
p_1\varphi_2^2+p_2\varphi_2^2+p_3\varphi_1^2 \leq \psi_2^2,
\end{eqnarray}
\begin{eqnarray}\label{App:B-CoC2-1st-3dim}
p_1\varphi_3^2+p_2\varphi_1^2+p_3\varphi_3^2 \leq \psi_3^2,
\end{eqnarray}
which gives us the CoC (for three-dimensional systems) defined by Eq.~\eqref{Condition2}. Furthermore, in the limit of orthonormal basis, equality holds in Eqs.~\eqref{App:B-CoC1-1st-3dim} and \eqref{App:B-CoC2-1st-3dim}, and also Eqs.~\eqref{App:B-CoC1-1st-3dim} and \eqref{App:B-CoC2-1st-3dim} become the same as Eqs.~\eqref{APP:B-3dim-1st-probabilities-2nd-eq} and \eqref{APP:B-3dim-1st-probabilities-3rd-eq}, respectively.
Since there are two subcases for three-dimensional systems, each of these steps can be repeated for the case $\tilde{\psi}_2 \leq \tilde{\varphi}_2$ of $d=3$.

This procedure can be easily generalized to the case of high-dimensional systems by following a similar path.
Using the Kraus operators defined by Eqs.~\eqref{Kraus-d-level1} and \eqref{Kraus-d-level2},
the completeness relation given by Eq.~\eqref{Completeness-KnFm} gives us the following $d(d+1)/2$ equations: $d$ equations in the form
\begin{eqnarray}\label{App-d-dim-eq1232d}
\sum_{i=1}^{2d} c_{j,i}^2 = 1,
\end{eqnarray}
for $j=1,2,\dots,d$; and $(d-1)d/2$ equations in the form
\begin{eqnarray}\label{App-d-dim-eq-rest}
\mu \sum_{i=1}^{d} c_{j,i} c_{l,i} + \sum_{i=d+1}^{d} c_{j,i} c_{l,i} = \mu,
\end{eqnarray}
for $j=1,2,\dots,(d-1)$, $l=2,3,\dots,d$, and $j<l$. The unitary transformations---permutations---presented in Ref.~\cite{Torun-DetCoherence}
give us the index functions. Also, from $F_m\ket{\psi} = 0 $ we have
\begin{eqnarray}\label{App:B-Kraus-Fm-ddim-condition}
c_{1,i} = -\big(c_{2,i}\psi_2 + c_{3,i}\psi_3 + \dots + c_{d,i}\psi_d\big)/\psi_1,
\end{eqnarray}
for $i=(d+1),\dots,2d$. Then, we divide the above $d(d+1)/2$ equations (Eqs.~\eqref{App-d-dim-eq1232d} and \eqref{App-d-dim-eq-rest})
into three separate groups and examine each one step by step. These are the final steps for obtaining the CoC given by Eq.~\eqref{Condition2}
and the majorization condition for superposition given by Eq.~\eqref{Maj-Superposition-Condition1}. With this formulation above, conditions for a class of superposition transformations and the transformation itself can be achieved effectively.


\begin{thebibliography}{65}%
\makeatletter
\providecommand \@ifxundefined [1]{%
 \@ifx{#1\undefined}
}%
\providecommand \@ifnum [1]{%
 \ifnum #1\expandafter \@firstoftwo
 \else \expandafter \@secondoftwo
 \fi
}%
\providecommand \@ifx [1]{%
 \ifx #1\expandafter \@firstoftwo
 \else \expandafter \@secondoftwo
 \fi
}%
\providecommand \natexlab [1]{#1}%
\providecommand \enquote  [1]{``#1''}%
\providecommand \bibnamefont  [1]{#1}%
\providecommand \bibfnamefont [1]{#1}%
\providecommand \citenamefont [1]{#1}%
\providecommand \href@noop [0]{\@secondoftwo}%
\providecommand \href [0]{\begingroup \@sanitize@url \@href}%
\providecommand \@href[1]{\@@startlink{#1}\@@href}%
\providecommand \@@href[1]{\endgroup#1\@@endlink}%
\providecommand \@sanitize@url [0]{\catcode `\\12\catcode `\$12\catcode
  `\&12\catcode `\#12\catcode `\^12\catcode `\_12\catcode `\%12\relax}%
\providecommand \@@startlink[1]{}%
\providecommand \@@endlink[0]{}%
\providecommand \url  [0]{\begingroup\@sanitize@url \@url }%
\providecommand \@url [1]{\endgroup\@href {#1}{\urlprefix }}%
\providecommand \urlprefix  [0]{URL }%
\providecommand \Eprint [0]{\href }%
\providecommand \doibase [0]{http://dx.doi.org/}%
\providecommand \selectlanguage [0]{\@gobble}%
\providecommand \bibinfo  [0]{\@secondoftwo}%
\providecommand \bibfield  [0]{\@secondoftwo}%
\providecommand \translation [1]{[#1]}%
\providecommand \BibitemOpen [0]{}%
\providecommand \bibitemStop [0]{}%
\providecommand \bibitemNoStop [0]{.\EOS\space}%
\providecommand \EOS [0]{\spacefactor3000\relax}%
\providecommand \BibitemShut  [1]{\csname bibitem#1\endcsname}%
\let\auto@bib@innerbib\@empty
\bibitem [{\citenamefont {Dirac}(1930)}]{Dirac-Superposition}%
  \BibitemOpen
  \bibfield  {author} {\bibinfo {author} {\bibfnamefont {P.~A.~M.}\
  \bibnamefont {Dirac}},\ }\href@noop {} {\emph {\bibinfo {title} {The
  Principles of Quantum Mechanics}}},\ \bibinfo {edition} {3rd}\ ed.\ (\bibinfo
   {publisher} {Clarendon Press, Oxford},\ \bibinfo {year} {1930})\BibitemShut
  {NoStop}%
\bibitem [{\citenamefont {Feix}\ \emph {et~al.}(2015)\citenamefont {Feix},
  \citenamefont {Ara\'ujo},\ and\ \citenamefont {Brukner}}]{Feix-CComp}%
  \BibitemOpen
  \bibfield  {author} {\bibinfo {author} {\bibfnamefont {A.}~\bibnamefont
  {Feix}}, \bibinfo {author} {\bibfnamefont {M.}~\bibnamefont {Ara\'ujo}}, \
  and\ \bibinfo {author} {\bibfnamefont {{\v{C}}.}~\bibnamefont {Brukner}},\
  }\href {\doibase 10.1103/PhysRevA.92.052326} {\bibfield  {journal} {\bibinfo
  {journal} {Phys. Rev. A}\ }\textbf {\bibinfo {volume} {92}},\ \bibinfo
  {pages} {052326} (\bibinfo {year} {2015})}\BibitemShut {NoStop}%
\bibitem [{\citenamefont {Gu\'erin}\ \emph {et~al.}(2016)\citenamefont
  {Gu\'erin}, \citenamefont {Feix}, \citenamefont {Ara\'ujo},\ and\
  \citenamefont {Brukner}}]{Brukner-Complexity}%
  \BibitemOpen
  \bibfield  {author} {\bibinfo {author} {\bibfnamefont {P.~A.}\ \bibnamefont
  {Gu\'erin}}, \bibinfo {author} {\bibfnamefont {A.}~\bibnamefont {Feix}},
  \bibinfo {author} {\bibfnamefont {M.}~\bibnamefont {Ara\'ujo}}, \ and\
  \bibinfo {author} {\bibfnamefont {{\v{C}}.}~\bibnamefont {Brukner}},\ }\href
  {\doibase 10.1103/PhysRevLett.117.100502} {\bibfield  {journal} {\bibinfo
  {journal} {Phys. Rev. Lett.}\ }\textbf {\bibinfo {volume} {117}},\ \bibinfo
  {pages} {100502} (\bibinfo {year} {2016})}\BibitemShut {NoStop}%
\bibitem [{\citenamefont {Theurer}\ \emph {et~al.}(2017)\citenamefont
  {Theurer}, \citenamefont {Killoran}, \citenamefont {Egloff},\ and\
  \citenamefont {Plenio}}]{Plenio-RTofS}%
  \BibitemOpen
  \bibfield  {author} {\bibinfo {author} {\bibfnamefont {T.}~\bibnamefont
  {Theurer}}, \bibinfo {author} {\bibfnamefont {N.}~\bibnamefont {Killoran}},
  \bibinfo {author} {\bibfnamefont {D.}~\bibnamefont {Egloff}}, \ and\ \bibinfo
  {author} {\bibfnamefont {M.~B.}\ \bibnamefont {Plenio}},\ }\href {\doibase
  10.1103/PhysRevLett.119.230401} {\bibfield  {journal} {\bibinfo  {journal}
  {Phys. Rev. Lett.}\ }\textbf {\bibinfo {volume} {119}},\ \bibinfo {pages}
  {230401} (\bibinfo {year} {2017})}\BibitemShut {NoStop}%
\bibitem [{\citenamefont {Chitambar}\ and\ \citenamefont
  {Gour}(2019)}]{Chitambar-QRTs}%
  \BibitemOpen
  \bibfield  {author} {\bibinfo {author} {\bibfnamefont {E.}~\bibnamefont
  {Chitambar}}\ and\ \bibinfo {author} {\bibfnamefont {G.}~\bibnamefont
  {Gour}},\ }\href {\doibase 10.1103/RevModPhys.91.025001} {\bibfield
  {journal} {\bibinfo  {journal} {Rev. Mod. Phys.}\ }\textbf {\bibinfo {volume}
  {91}},\ \bibinfo {pages} {025001} (\bibinfo {year} {2019})}\BibitemShut
  {NoStop}%
\bibitem [{\citenamefont {Takagi}\ and\ \citenamefont
  {Regula}(2019)}]{Regula-QRTs}%
  \BibitemOpen
  \bibfield  {author} {\bibinfo {author} {\bibfnamefont {R.}~\bibnamefont
  {Takagi}}\ and\ \bibinfo {author} {\bibfnamefont {B.}~\bibnamefont
  {Regula}},\ }\href {\doibase 10.1103/PhysRevX.9.031053} {\bibfield  {journal}
  {\bibinfo  {journal} {Phys. Rev. X}\ }\textbf {\bibinfo {volume} {9}},\
  \bibinfo {pages} {031053} (\bibinfo {year} {2019})}\BibitemShut {NoStop}%
\bibitem [{\citenamefont {Horodecki}\ \emph {et~al.}(2009)\citenamefont
  {Horodecki}, \citenamefont {Horodecki}, \citenamefont {Horodecki},\ and\
  \citenamefont {Horodecki}}]{Horodecki-QE}%
  \BibitemOpen
  \bibfield  {author} {\bibinfo {author} {\bibfnamefont {R.}~\bibnamefont
  {Horodecki}}, \bibinfo {author} {\bibfnamefont {P.}~\bibnamefont
  {Horodecki}}, \bibinfo {author} {\bibfnamefont {M.}~\bibnamefont
  {Horodecki}}, \ and\ \bibinfo {author} {\bibfnamefont {K.}~\bibnamefont
  {Horodecki}},\ }\href {\doibase 10.1103/RevModPhys.81.865} {\bibfield
  {journal} {\bibinfo  {journal} {Rev. Mod. Phys.}\ }\textbf {\bibinfo {volume}
  {81}},\ \bibinfo {pages} {865} (\bibinfo {year} {2009})}\BibitemShut
  {NoStop}%
\bibitem [{\citenamefont {Sparaciari}\ \emph {et~al.}(2020)\citenamefont
  {Sparaciari}, \citenamefont {del Rio}, \citenamefont {Scandolo},
  \citenamefont {Faist},\ and\ \citenamefont
  {Oppenheim}}]{Sparaciari2020firstlawofgeneral}%
  \BibitemOpen
  \bibfield  {author} {\bibinfo {author} {\bibfnamefont {C.}~\bibnamefont
  {Sparaciari}}, \bibinfo {author} {\bibfnamefont {L.}~\bibnamefont {del Rio}},
  \bibinfo {author} {\bibfnamefont {C.~M.}\ \bibnamefont {Scandolo}}, \bibinfo
  {author} {\bibfnamefont {P.}~\bibnamefont {Faist}}, \ and\ \bibinfo {author}
  {\bibfnamefont {J.}~\bibnamefont {Oppenheim}},\ }\href {\doibase
  10.22331/q-2020-04-30-259} {\bibfield  {journal} {\bibinfo  {journal}
  {{Quantum}}\ }\textbf {\bibinfo {volume} {4}},\ \bibinfo {pages} {259}
  (\bibinfo {year} {2020})}\BibitemShut {NoStop}%
\bibitem [{\citenamefont {Horodecki}\ and\ \citenamefont
  {Oppenheim}(2013)}]{Horodecki-QRT}%
  \BibitemOpen
  \bibfield  {author} {\bibinfo {author} {\bibfnamefont {M.}~\bibnamefont
  {Horodecki}}\ and\ \bibinfo {author} {\bibfnamefont {J.}~\bibnamefont
  {Oppenheim}},\ }\href {\doibase 10.1142/S0217979213450197} {\bibfield
  {journal} {\bibinfo  {journal} {Int. J. Mod. Phys. B}\ }\textbf {\bibinfo
  {volume} {27}},\ \bibinfo {pages} {1345019} (\bibinfo {year}
  {2013})}\BibitemShut {NoStop}%
\bibitem [{\citenamefont {Veitch}\ \emph {et~al.}(2014)\citenamefont {Veitch},
  \citenamefont {Mousavian}, \citenamefont {Gottesman},\ and\ \citenamefont
  {Emerson}}]{Veitch_2014}%
  \BibitemOpen
  \bibfield  {author} {\bibinfo {author} {\bibfnamefont {V.}~\bibnamefont
  {Veitch}}, \bibinfo {author} {\bibfnamefont {S.~A.~H.}\ \bibnamefont
  {Mousavian}}, \bibinfo {author} {\bibfnamefont {D.}~\bibnamefont
  {Gottesman}}, \ and\ \bibinfo {author} {\bibfnamefont {J.}~\bibnamefont
  {Emerson}},\ }\href {\doibase 10.1088/1367-2630/16/1/013009} {\bibfield
  {journal} {\bibinfo  {journal} {N. J. Phys.}\ }\textbf {\bibinfo {volume}
  {16}},\ \bibinfo {pages} {013009} (\bibinfo {year} {2014})}\BibitemShut
  {NoStop}%
\bibitem [{\citenamefont {del Rio}\ \emph {et~al.}(2015)\citenamefont {del
  Rio}, \citenamefont {Kraemer},\ and\ \citenamefont
  {Renner}}]{rio2015resource}%
  \BibitemOpen
  \bibfield  {author} {\bibinfo {author} {\bibfnamefont {L.}~\bibnamefont {del
  Rio}}, \bibinfo {author} {\bibfnamefont {L.}~\bibnamefont {Kraemer}}, \ and\
  \bibinfo {author} {\bibfnamefont {R.}~\bibnamefont {Renner}},\ }\href@noop {}
  {} (\bibinfo {year} {2015}),\ \Eprint {http://arxiv.org/abs/1511.08818}
  {arXiv:1511.08818} \BibitemShut {NoStop}%
\bibitem [{\citenamefont {Brand\~ao}\ and\ \citenamefont
  {Gour}(2015)}]{Brandao-QRT-Reversible}%
  \BibitemOpen
  \bibfield  {author} {\bibinfo {author} {\bibfnamefont {F.~G. S.~L.}\
  \bibnamefont {Brand\~ao}}\ and\ \bibinfo {author} {\bibfnamefont
  {G.}~\bibnamefont {Gour}},\ }\href {\doibase 10.1103/PhysRevLett.115.070503}
  {\bibfield  {journal} {\bibinfo  {journal} {Phys. Rev. Lett.}\ }\textbf
  {\bibinfo {volume} {115}},\ \bibinfo {pages} {070503} (\bibinfo {year}
  {2015})}\BibitemShut {NoStop}%
\bibitem [{\citenamefont {de~Vicente}\ and\ \citenamefont
  {Streltsov}(2016)}]{de_Vicente_2016}%
  \BibitemOpen
  \bibfield  {author} {\bibinfo {author} {\bibfnamefont {J.~I.}\ \bibnamefont
  {de~Vicente}}\ and\ \bibinfo {author} {\bibfnamefont {A.}~\bibnamefont
  {Streltsov}},\ }\href {\doibase 10.1088/1751-8121/50/4/045301} {\bibfield
  {journal} {\bibinfo  {journal} {J. Phys. A: Math. Theor.}\ }\textbf {\bibinfo
  {volume} {50}},\ \bibinfo {pages} {045301} (\bibinfo {year}
  {2016})}\BibitemShut {NoStop}%
\bibitem [{\citenamefont {Sparaciari}\ \emph {et~al.}(2017)\citenamefont
  {Sparaciari}, \citenamefont {Oppenheim},\ and\ \citenamefont
  {Fritz}}]{RT-Heat-and-Work}%
  \BibitemOpen
  \bibfield  {author} {\bibinfo {author} {\bibfnamefont {C.}~\bibnamefont
  {Sparaciari}}, \bibinfo {author} {\bibfnamefont {J.}~\bibnamefont
  {Oppenheim}}, \ and\ \bibinfo {author} {\bibfnamefont {T.}~\bibnamefont
  {Fritz}},\ }\href {\doibase 10.1103/PhysRevA.96.052112} {\bibfield  {journal}
  {\bibinfo  {journal} {Phys. Rev. A}\ }\textbf {\bibinfo {volume} {96}},\
  \bibinfo {pages} {052112} (\bibinfo {year} {2017})}\BibitemShut {NoStop}%
\bibitem [{\citenamefont {Gour}(2017)}]{Gour-QRT-SingleShot}%
  \BibitemOpen
  \bibfield  {author} {\bibinfo {author} {\bibfnamefont {G.}~\bibnamefont
  {Gour}},\ }\href {\doibase 10.1103/PhysRevA.95.062314} {\bibfield  {journal}
  {\bibinfo  {journal} {Phys. Rev. A}\ }\textbf {\bibinfo {volume} {95}},\
  \bibinfo {pages} {062314} (\bibinfo {year} {2017})}\BibitemShut {NoStop}%
\bibitem [{\citenamefont {Liu}\ \emph {et~al.}(2017)\citenamefont {Liu},
  \citenamefont {Hu},\ and\ \citenamefont
  {Lloyd}}]{Liu-ResourceDestroyingMaps}%
  \BibitemOpen
  \bibfield  {author} {\bibinfo {author} {\bibfnamefont {Z.-W.}\ \bibnamefont
  {Liu}}, \bibinfo {author} {\bibfnamefont {X.}~\bibnamefont {Hu}}, \ and\
  \bibinfo {author} {\bibfnamefont {S.}~\bibnamefont {Lloyd}},\ }\href
  {\doibase 10.1103/PhysRevLett.118.060502} {\bibfield  {journal} {\bibinfo
  {journal} {Phys. Rev. Lett.}\ }\textbf {\bibinfo {volume} {118}},\ \bibinfo
  {pages} {060502} (\bibinfo {year} {2017})}\BibitemShut {NoStop}%
\bibitem [{\citenamefont {Howard}\ and\ \citenamefont
  {Campbell}(2017)}]{Howard-RTMagic}%
  \BibitemOpen
  \bibfield  {author} {\bibinfo {author} {\bibfnamefont {M.}~\bibnamefont
  {Howard}}\ and\ \bibinfo {author} {\bibfnamefont {E.}~\bibnamefont
  {Campbell}},\ }\href {\doibase 10.1103/PhysRevLett.118.090501} {\bibfield
  {journal} {\bibinfo  {journal} {Phys. Rev. Lett.}\ }\textbf {\bibinfo
  {volume} {118}},\ \bibinfo {pages} {090501} (\bibinfo {year}
  {2017})}\BibitemShut {NoStop}%
\bibitem [{\citenamefont {Lami}\ \emph {et~al.}(2018)\citenamefont {Lami},
  \citenamefont {Regula}, \citenamefont {Wang}, \citenamefont {Nichols},
  \citenamefont {Winter},\ and\ \citenamefont {Adesso}}]{Gaussin-QRT}%
  \BibitemOpen
  \bibfield  {author} {\bibinfo {author} {\bibfnamefont {L.}~\bibnamefont
  {Lami}}, \bibinfo {author} {\bibfnamefont {B.}~\bibnamefont {Regula}},
  \bibinfo {author} {\bibfnamefont {X.}~\bibnamefont {Wang}}, \bibinfo {author}
  {\bibfnamefont {R.}~\bibnamefont {Nichols}}, \bibinfo {author} {\bibfnamefont
  {A.}~\bibnamefont {Winter}}, \ and\ \bibinfo {author} {\bibfnamefont
  {G.}~\bibnamefont {Adesso}},\ }\href {\doibase 10.1103/PhysRevA.98.022335}
  {\bibfield  {journal} {\bibinfo  {journal} {Phys. Rev. A}\ }\textbf {\bibinfo
  {volume} {98}},\ \bibinfo {pages} {022335} (\bibinfo {year}
  {2018})}\BibitemShut {NoStop}%
\bibitem [{\citenamefont {Takagi}\ \emph {et~al.}(2019)\citenamefont {Takagi},
  \citenamefont {Regula}, \citenamefont {Bu}, \citenamefont {Liu},\ and\
  \citenamefont {Adesso}}]{Takagi_OpAdvQR}%
  \BibitemOpen
  \bibfield  {author} {\bibinfo {author} {\bibfnamefont {R.}~\bibnamefont
  {Takagi}}, \bibinfo {author} {\bibfnamefont {B.}~\bibnamefont {Regula}},
  \bibinfo {author} {\bibfnamefont {K.}~\bibnamefont {Bu}}, \bibinfo {author}
  {\bibfnamefont {Z.-W.}\ \bibnamefont {Liu}}, \ and\ \bibinfo {author}
  {\bibfnamefont {G.}~\bibnamefont {Adesso}},\ }\href {\doibase
  10.1103/PhysRevLett.122.140402} {\bibfield  {journal} {\bibinfo  {journal}
  {Phys. Rev. Lett.}\ }\textbf {\bibinfo {volume} {122}},\ \bibinfo {pages}
  {140402} (\bibinfo {year} {2019})}\BibitemShut {NoStop}%
\bibitem [{\citenamefont {Oszmaniec}\ and\ \citenamefont
  {Biswas}(2019)}]{Oszmaniec2019operational}%
  \BibitemOpen
  \bibfield  {author} {\bibinfo {author} {\bibfnamefont {M.}~\bibnamefont
  {Oszmaniec}}\ and\ \bibinfo {author} {\bibfnamefont {T.}~\bibnamefont
  {Biswas}},\ }\href {\doibase 10.22331/q-2019-04-26-133} {\bibfield  {journal}
  {\bibinfo  {journal} {{Quantum}}\ }\textbf {\bibinfo {volume} {3}},\ \bibinfo
  {pages} {133} (\bibinfo {year} {2019})}\BibitemShut {NoStop}%
\bibitem [{\citenamefont {Liu}\ and\ \citenamefont
  {Winter}(2019)}]{liu2019resource}%
  \BibitemOpen
  \bibfield  {author} {\bibinfo {author} {\bibfnamefont {Z.-W.}\ \bibnamefont
  {Liu}}\ and\ \bibinfo {author} {\bibfnamefont {A.}~\bibnamefont {Winter}},\
  }\href@noop {} {} (\bibinfo {year} {2019}),\ \Eprint
  {http://arxiv.org/abs/1904.04201} {arXiv:1904.04201} \BibitemShut {NoStop}%
\bibitem [{\citenamefont {Wang}\ and\ \citenamefont
  {Wilde}(2019)}]{Wang-RTAsymetric}%
  \BibitemOpen
  \bibfield  {author} {\bibinfo {author} {\bibfnamefont {X.}~\bibnamefont
  {Wang}}\ and\ \bibinfo {author} {\bibfnamefont {M.~M.}\ \bibnamefont
  {Wilde}},\ }\href {\doibase 10.1103/PhysRevResearch.1.033170} {\bibfield
  {journal} {\bibinfo  {journal} {Phys. Rev. Research}\ }\textbf {\bibinfo
  {volume} {1}},\ \bibinfo {pages} {033170} (\bibinfo {year}
  {2019})}\BibitemShut {NoStop}%
\bibitem [{\citenamefont {Li}\ \emph {et~al.}(2020)\citenamefont {Li},
  \citenamefont {Bu},\ and\ \citenamefont {Liu}}]{Li-QuantifyingRCQC}%
  \BibitemOpen
  \bibfield  {author} {\bibinfo {author} {\bibfnamefont {L.}~\bibnamefont
  {Li}}, \bibinfo {author} {\bibfnamefont {K.}~\bibnamefont {Bu}}, \ and\
  \bibinfo {author} {\bibfnamefont {Z.-W.}\ \bibnamefont {Liu}},\ }\href
  {\doibase 10.1103/PhysRevA.101.022335} {\bibfield  {journal} {\bibinfo
  {journal} {Phys. Rev. A}\ }\textbf {\bibinfo {volume} {101}},\ \bibinfo
  {pages} {022335} (\bibinfo {year} {2020})}\BibitemShut {NoStop}%
\bibitem [{\citenamefont {Buscemi}\ \emph {et~al.}(2020)\citenamefont
  {Buscemi}, \citenamefont {Chitambar},\ and\ \citenamefont
  {Zhou}}]{CRT-Incompatibility}%
  \BibitemOpen
  \bibfield  {author} {\bibinfo {author} {\bibfnamefont {F.}~\bibnamefont
  {Buscemi}}, \bibinfo {author} {\bibfnamefont {E.}~\bibnamefont {Chitambar}},
  \ and\ \bibinfo {author} {\bibfnamefont {W.}~\bibnamefont {Zhou}},\ }\href
  {\doibase 10.1103/PhysRevLett.124.120401} {\bibfield  {journal} {\bibinfo
  {journal} {Phys. Rev. Lett.}\ }\textbf {\bibinfo {volume} {124}},\ \bibinfo
  {pages} {120401} (\bibinfo {year} {2020})}\BibitemShut {NoStop}%
\bibitem [{\citenamefont {Kristj\'{a}nsson}\ \emph {et~al.}(2019)\citenamefont
  {Kristj\'{a}nsson}, \citenamefont {Chiribella}, \citenamefont {Salek},
  \citenamefont {Ebler},\ and\ \citenamefont
  {Wilson}}]{kristjnsson2019resource}%
  \BibitemOpen
  \bibfield  {author} {\bibinfo {author} {\bibfnamefont {H.}~\bibnamefont
  {Kristj\'{a}nsson}}, \bibinfo {author} {\bibfnamefont {G.}~\bibnamefont
  {Chiribella}}, \bibinfo {author} {\bibfnamefont {S.}~\bibnamefont {Salek}},
  \bibinfo {author} {\bibfnamefont {D.}~\bibnamefont {Ebler}}, \ and\ \bibinfo
  {author} {\bibfnamefont {M.}~\bibnamefont {Wilson}},\ }\href@noop {} {}
  (\bibinfo {year} {2019}),\ \Eprint {http://arxiv.org/abs/1910.08197}
  {arXiv:1910.08197} \BibitemShut {NoStop}%
\bibitem [{\citenamefont {Liu}\ and\ \citenamefont {Yuan}(2020)}]{RT-of-QC}%
  \BibitemOpen
  \bibfield  {author} {\bibinfo {author} {\bibfnamefont {Y.}~\bibnamefont
  {Liu}}\ and\ \bibinfo {author} {\bibfnamefont {X.}~\bibnamefont {Yuan}},\
  }\href {\doibase 10.1103/PhysRevResearch.2.012035} {\bibfield  {journal}
  {\bibinfo  {journal} {Phys. Rev. Research}\ }\textbf {\bibinfo {volume}
  {2}},\ \bibinfo {pages} {012035} (\bibinfo {year} {2020})}\BibitemShut
  {NoStop}%
\bibitem [{\citenamefont {Wolfe}\ \emph {et~al.}(2020)\citenamefont {Wolfe},
  \citenamefont {Schmid}, \citenamefont {Sainz}, \citenamefont {Kunjwal},\ and\
  \citenamefont {Spekkens}}]{Wolfe2020quantifyingbell}%
  \BibitemOpen
  \bibfield  {author} {\bibinfo {author} {\bibfnamefont {E.}~\bibnamefont
  {Wolfe}}, \bibinfo {author} {\bibfnamefont {D.}~\bibnamefont {Schmid}},
  \bibinfo {author} {\bibfnamefont {A.~B.}\ \bibnamefont {Sainz}}, \bibinfo
  {author} {\bibfnamefont {R.}~\bibnamefont {Kunjwal}}, \ and\ \bibinfo
  {author} {\bibfnamefont {R.~W.}\ \bibnamefont {Spekkens}},\ }\href {\doibase
  10.22331/q-2020-06-08-280} {\bibfield  {journal} {\bibinfo  {journal}
  {{Quantum}}\ }\textbf {\bibinfo {volume} {4}},\ \bibinfo {pages} {280}
  (\bibinfo {year} {2020})}\BibitemShut {NoStop}%
\bibitem [{\citenamefont {Aberg}(2006)}]{Aberg-Superposition}%
  \BibitemOpen
  \bibfield  {author} {\bibinfo {author} {\bibfnamefont {J.}~\bibnamefont
  {Aberg}},\ }\href@noop {} {} (\bibinfo {year} {2006}),\ \Eprint
  {http://arxiv.org/abs/quant-ph/0612146} {arXiv:quant-ph/0612146} \BibitemShut
  {NoStop}%
\bibitem [{\citenamefont {Das}\ \emph {et~al.}(2020)\citenamefont {Das},
  \citenamefont {Mukhopadhyay}, \citenamefont {Roy}, \citenamefont
  {Bhattacharya}, \citenamefont {Sen(De)},\ and\ \citenamefont
  {Sen}}]{Das-NonorthogonalCoherence}%
  \BibitemOpen
  \bibfield  {author} {\bibinfo {author} {\bibfnamefont {S.}~\bibnamefont
  {Das}}, \bibinfo {author} {\bibfnamefont {C.}~\bibnamefont {Mukhopadhyay}},
  \bibinfo {author} {\bibfnamefont {S.~S.}\ \bibnamefont {Roy}}, \bibinfo
  {author} {\bibfnamefont {S.}~\bibnamefont {Bhattacharya}}, \bibinfo {author}
  {\bibfnamefont {A.}~\bibnamefont {Sen(De)}}, \ and\ \bibinfo {author}
  {\bibfnamefont {U.}~\bibnamefont {Sen}},\ }\href {\doibase
  10.1088/1751-8121/ab741f} {\bibfield  {journal} {\bibinfo  {journal} {J.
  Phys. A: Math. Theor.}\ }\textbf {\bibinfo {volume} {53}},\ \bibinfo {pages}
  {115301} (\bibinfo {year} {2020})}\BibitemShut {NoStop}%
\bibitem [{\citenamefont {Baumgratz}\ \emph {et~al.}(2014)\citenamefont
  {Baumgratz}, \citenamefont {Cramer},\ and\ \citenamefont
  {Plenio}}]{Baumgratz-Coherence}%
  \BibitemOpen
  \bibfield  {author} {\bibinfo {author} {\bibfnamefont {T.}~\bibnamefont
  {Baumgratz}}, \bibinfo {author} {\bibfnamefont {M.}~\bibnamefont {Cramer}}, \
  and\ \bibinfo {author} {\bibfnamefont {M.~B.}\ \bibnamefont {Plenio}},\
  }\href {\doibase 10.1103/PhysRevLett.113.140401} {\bibfield  {journal}
  {\bibinfo  {journal} {Phys. Rev. Lett.}\ }\textbf {\bibinfo {volume} {113}},\
  \bibinfo {pages} {140401} (\bibinfo {year} {2014})}\BibitemShut {NoStop}%
\bibitem [{\citenamefont {Bera}(2019)}]{Bera-QS}%
  \BibitemOpen
  \bibfield  {author} {\bibinfo {author} {\bibfnamefont {M.~N.}\ \bibnamefont
  {Bera}},\ }\href {\doibase 10.1103/PhysRevA.100.042307} {\bibfield  {journal}
  {\bibinfo  {journal} {Phys. Rev. A}\ }\textbf {\bibinfo {volume} {100}},\
  \bibinfo {pages} {042307} (\bibinfo {year} {2019})}\BibitemShut {NoStop}%
\bibitem [{\citenamefont {Nielsen}(1999)}]{Nielsen-Maj}%
  \BibitemOpen
  \bibfield  {author} {\bibinfo {author} {\bibfnamefont {M.~A.}\ \bibnamefont
  {Nielsen}},\ }\href {\doibase 10.1103/PhysRevLett.83.436} {\bibfield
  {journal} {\bibinfo  {journal} {Phys. Rev. Lett.}\ }\textbf {\bibinfo
  {volume} {83}},\ \bibinfo {pages} {436} (\bibinfo {year} {1999})}\BibitemShut
  {NoStop}%
\bibitem [{\citenamefont {Chefles}\ \emph {et~al.}(2004)\citenamefont
  {Chefles}, \citenamefont {Jozsa},\ and\ \citenamefont
  {Winter}}]{AWinter_2004}%
  \BibitemOpen
  \bibfield  {author} {\bibinfo {author} {\bibfnamefont {A.}~\bibnamefont
  {Chefles}}, \bibinfo {author} {\bibfnamefont {R.}~\bibnamefont {Jozsa}}, \
  and\ \bibinfo {author} {\bibfnamefont {A.}~\bibnamefont {Winter}},\ }\href
  {\doibase 10.1142/S0219749904000031} {\bibfield  {journal} {\bibinfo
  {journal} {Int. J. Quant. Inf.}\ }\textbf {\bibinfo {volume} {02}},\ \bibinfo
  {pages} {11} (\bibinfo {year} {2004})}\BibitemShut {NoStop}%
\bibitem [{\citenamefont {Bravyi}\ and\ \citenamefont
  {Kitaev}(2005)}]{Bravyi-MagicStatePuri}%
  \BibitemOpen
  \bibfield  {author} {\bibinfo {author} {\bibfnamefont {S.}~\bibnamefont
  {Bravyi}}\ and\ \bibinfo {author} {\bibfnamefont {A.}~\bibnamefont
  {Kitaev}},\ }\href {\doibase 10.1103/PhysRevA.71.022316} {\bibfield
  {journal} {\bibinfo  {journal} {Phys. Rev. A}\ }\textbf {\bibinfo {volume}
  {71}},\ \bibinfo {pages} {022316} (\bibinfo {year} {2005})}\BibitemShut
  {NoStop}%
\bibitem [{\citenamefont {{Brandao}}\ and\ \citenamefont
  {{Datta}}(2011)}]{Brandao-EntManip}%
  \BibitemOpen
  \bibfield  {author} {\bibinfo {author} {\bibfnamefont {F.~G. S.~L.}\
  \bibnamefont {{Brandao}}}\ and\ \bibinfo {author} {\bibfnamefont
  {N.}~\bibnamefont {{Datta}}},\ }\href {\doibase 10.1109/TIT.2011.2104531}
  {\bibfield  {journal} {\bibinfo  {journal} {IEEE Trans. Inf. Theory}\
  }\textbf {\bibinfo {volume} {57}},\ \bibinfo {pages} {1754} (\bibinfo {year}
  {2011})}\BibitemShut {NoStop}%
\bibitem [{\citenamefont {Marvian}\ and\ \citenamefont
  {Spekkens}(2013)}]{Marvian_2013}%
  \BibitemOpen
  \bibfield  {author} {\bibinfo {author} {\bibfnamefont {I.}~\bibnamefont
  {Marvian}}\ and\ \bibinfo {author} {\bibfnamefont {R.~W.}\ \bibnamefont
  {Spekkens}},\ }\href {\doibase 10.1088/1367-2630/15/3/033001} {\bibfield
  {journal} {\bibinfo  {journal} {N. J. Phys.}\ }\textbf {\bibinfo {volume}
  {15}},\ \bibinfo {pages} {033001} (\bibinfo {year} {2013})}\BibitemShut
  {NoStop}%
\bibitem [{\citenamefont {de~Vicente}\ \emph {et~al.}(2017)\citenamefont
  {de~Vicente}, \citenamefont {Spee}, \citenamefont {Sauerwein},\ and\
  \citenamefont {Kraus}}]{Vicente_EntManipulation}%
  \BibitemOpen
  \bibfield  {author} {\bibinfo {author} {\bibfnamefont {J.~I.}\ \bibnamefont
  {de~Vicente}}, \bibinfo {author} {\bibfnamefont {C.}~\bibnamefont {Spee}},
  \bibinfo {author} {\bibfnamefont {D.}~\bibnamefont {Sauerwein}}, \ and\
  \bibinfo {author} {\bibfnamefont {B.}~\bibnamefont {Kraus}},\ }\href
  {\doibase 10.1103/PhysRevA.95.012323} {\bibfield  {journal} {\bibinfo
  {journal} {Phys. Rev. A}\ }\textbf {\bibinfo {volume} {95}},\ \bibinfo
  {pages} {012323} (\bibinfo {year} {2017})}\BibitemShut {NoStop}%
\bibitem [{\citenamefont {Wang}\ and\ \citenamefont
  {Duan}(2017)}]{Wang_AsymptoticEntM}%
  \BibitemOpen
  \bibfield  {author} {\bibinfo {author} {\bibfnamefont {X.}~\bibnamefont
  {Wang}}\ and\ \bibinfo {author} {\bibfnamefont {R.}~\bibnamefont {Duan}},\
  }\href {\doibase 10.1103/PhysRevLett.119.180506} {\bibfield  {journal}
  {\bibinfo  {journal} {Phys. Rev. Lett.}\ }\textbf {\bibinfo {volume} {119}},\
  \bibinfo {pages} {180506} (\bibinfo {year} {2017})}\BibitemShut {NoStop}%
\bibitem [{\citenamefont {Du}\ \emph {et~al.}(2015)\citenamefont {Du},
  \citenamefont {Bai},\ and\ \citenamefont {Guo}}]{Du-CoherenceMaj}%
  \BibitemOpen
  \bibfield  {author} {\bibinfo {author} {\bibfnamefont {S.}~\bibnamefont
  {Du}}, \bibinfo {author} {\bibfnamefont {Z.}~\bibnamefont {Bai}}, \ and\
  \bibinfo {author} {\bibfnamefont {Y.}~\bibnamefont {Guo}},\ }\href {\doibase
  10.1103/PhysRevA.91.052120} {\bibfield  {journal} {\bibinfo  {journal} {Phys.
  Rev. A}\ }\textbf {\bibinfo {volume} {91}},\ \bibinfo {pages} {052120}
  (\bibinfo {year} {2015})}\BibitemShut {NoStop}%
\bibitem [{\citenamefont {Liu}\ \emph {et~al.}(2019)\citenamefont {Liu},
  \citenamefont {Bu},\ and\ \citenamefont {Takagi}}]{Liu-OneShotRT}%
  \BibitemOpen
  \bibfield  {author} {\bibinfo {author} {\bibfnamefont {Z.-W.}\ \bibnamefont
  {Liu}}, \bibinfo {author} {\bibfnamefont {K.}~\bibnamefont {Bu}}, \ and\
  \bibinfo {author} {\bibfnamefont {R.}~\bibnamefont {Takagi}},\ }\href
  {\doibase 10.1103/PhysRevLett.123.020401} {\bibfield  {journal} {\bibinfo
  {journal} {Phys. Rev. Lett.}\ }\textbf {\bibinfo {volume} {123}},\ \bibinfo
  {pages} {020401} (\bibinfo {year} {2019})}\BibitemShut {NoStop}%
\bibitem [{\citenamefont {Regula}\ \emph
  {et~al.}(2020{\natexlab{a}})\citenamefont {Regula}, \citenamefont {Bu},
  \citenamefont {Takagi},\ and\ \citenamefont {Liu}}]{Regula-OneShotDistill}%
  \BibitemOpen
  \bibfield  {author} {\bibinfo {author} {\bibfnamefont {B.}~\bibnamefont
  {Regula}}, \bibinfo {author} {\bibfnamefont {K.}~\bibnamefont {Bu}}, \bibinfo
  {author} {\bibfnamefont {R.}~\bibnamefont {Takagi}}, \ and\ \bibinfo {author}
  {\bibfnamefont {Z.-W.}\ \bibnamefont {Liu}},\ }\href {\doibase
  10.1103/PhysRevA.101.062315} {\bibfield  {journal} {\bibinfo  {journal}
  {Phys. Rev. A}\ }\textbf {\bibinfo {volume} {101}},\ \bibinfo {pages}
  {062315} (\bibinfo {year} {2020}{\natexlab{a}})}\BibitemShut {NoStop}%
\bibitem [{\citenamefont {Torun}\ \emph {et~al.}(2019)\citenamefont {Torun},
  \citenamefont {Lami}, \citenamefont {Adesso},\ and\ \citenamefont
  {Yildiz}}]{Torun-CoherenceDistill}%
  \BibitemOpen
  \bibfield  {author} {\bibinfo {author} {\bibfnamefont {G.}~\bibnamefont
  {Torun}}, \bibinfo {author} {\bibfnamefont {L.}~\bibnamefont {Lami}},
  \bibinfo {author} {\bibfnamefont {G.}~\bibnamefont {Adesso}}, \ and\ \bibinfo
  {author} {\bibfnamefont {A.}~\bibnamefont {Yildiz}},\ }\href {\doibase
  10.1103/PhysRevA.99.012321} {\bibfield  {journal} {\bibinfo  {journal} {Phys.
  Rev. A}\ }\textbf {\bibinfo {volume} {99}},\ \bibinfo {pages} {012321}
  (\bibinfo {year} {2019})}\BibitemShut {NoStop}%
\bibitem [{\citenamefont {Regula}\ \emph
  {et~al.}(2020{\natexlab{b}})\citenamefont {Regula}, \citenamefont
  {Narasimhachar}, \citenamefont {Buscemi},\ and\ \citenamefont
  {Gu}}]{Regula_DepCovOper}%
  \BibitemOpen
  \bibfield  {author} {\bibinfo {author} {\bibfnamefont {B.}~\bibnamefont
  {Regula}}, \bibinfo {author} {\bibfnamefont {V.}~\bibnamefont
  {Narasimhachar}}, \bibinfo {author} {\bibfnamefont {F.}~\bibnamefont
  {Buscemi}}, \ and\ \bibinfo {author} {\bibfnamefont {M.}~\bibnamefont {Gu}},\
  }\href {\doibase 10.1103/PhysRevResearch.2.013109} {\bibfield  {journal}
  {\bibinfo  {journal} {Phys. Rev. Research}\ }\textbf {\bibinfo {volume}
  {2}},\ \bibinfo {pages} {013109} (\bibinfo {year}
  {2020}{\natexlab{b}})}\BibitemShut {NoStop}%
\bibitem [{\citenamefont {{Fang}}\ \emph {et~al.}(2019)\citenamefont {{Fang}},
  \citenamefont {{Wang}}, \citenamefont {{Tomamichel}},\ and\ \citenamefont
  {{Duan}}}]{Fang-EntDistill}%
  \BibitemOpen
  \bibfield  {author} {\bibinfo {author} {\bibfnamefont {K.}~\bibnamefont
  {{Fang}}}, \bibinfo {author} {\bibfnamefont {X.}~\bibnamefont {{Wang}}},
  \bibinfo {author} {\bibfnamefont {M.}~\bibnamefont {{Tomamichel}}}, \ and\
  \bibinfo {author} {\bibfnamefont {R.}~\bibnamefont {{Duan}}},\ }\href
  {\doibase 10.1109/TIT.2019.2914688} {\bibfield  {journal} {\bibinfo
  {journal} {IEEE Trans. Inf. Theory}\ }\textbf {\bibinfo {volume} {65}},\
  \bibinfo {pages} {6454} (\bibinfo {year} {2019})}\BibitemShut {NoStop}%
\bibitem [{\citenamefont {Litinski}(2019)}]{Litinski2019magicstate}%
  \BibitemOpen
  \bibfield  {author} {\bibinfo {author} {\bibfnamefont {D.}~\bibnamefont
  {Litinski}},\ }\href {\doibase 10.22331/q-2019-12-02-205} {\bibfield
  {journal} {\bibinfo  {journal} {{Quantum}}\ }\textbf {\bibinfo {volume}
  {3}},\ \bibinfo {pages} {205} (\bibinfo {year} {2019})}\BibitemShut {NoStop}%
\bibitem [{\citenamefont {{Zhao}}\ \emph {et~al.}(2019)\citenamefont {{Zhao}},
  \citenamefont {{Liu}}, \citenamefont {{Yuan}}, \citenamefont {{Chitambar}},\
  and\ \citenamefont {{Winter}}}]{Zhao-CoherenceDistill}%
  \BibitemOpen
  \bibfield  {author} {\bibinfo {author} {\bibfnamefont {Q.}~\bibnamefont
  {{Zhao}}}, \bibinfo {author} {\bibfnamefont {Y.}~\bibnamefont {{Liu}}},
  \bibinfo {author} {\bibfnamefont {X.}~\bibnamefont {{Yuan}}}, \bibinfo
  {author} {\bibfnamefont {E.}~\bibnamefont {{Chitambar}}}, \ and\ \bibinfo
  {author} {\bibfnamefont {A.}~\bibnamefont {{Winter}}},\ }\href {\doibase
  10.1109/TIT.2019.2911102} {\bibfield  {journal} {\bibinfo  {journal} {IEEE
  Trans. Inf. Theory}\ }\textbf {\bibinfo {volume} {65}},\ \bibinfo {pages}
  {6441} (\bibinfo {year} {2019})}\BibitemShut {NoStop}%
\bibitem [{\citenamefont {Chitambar}\ \emph {et~al.}(2020)\citenamefont
  {Chitambar}, \citenamefont {de~Vicente}, \citenamefont {Girard},\ and\
  \citenamefont {Gour}}]{Chitambar_entManipulation}%
  \BibitemOpen
  \bibfield  {author} {\bibinfo {author} {\bibfnamefont {E.}~\bibnamefont
  {Chitambar}}, \bibinfo {author} {\bibfnamefont {J.~I.}\ \bibnamefont
  {de~Vicente}}, \bibinfo {author} {\bibfnamefont {M.~W.}\ \bibnamefont
  {Girard}}, \ and\ \bibinfo {author} {\bibfnamefont {G.}~\bibnamefont
  {Gour}},\ }\href {\doibase 10.1063/1.5124109} {\bibfield  {journal} {\bibinfo
   {journal} {J. Math. Phys.}\ }\textbf {\bibinfo {volume} {61}},\ \bibinfo
  {pages} {042201} (\bibinfo {year} {2020})}\BibitemShut {NoStop}%
\bibitem [{\citenamefont {Fang}\ and\ \citenamefont {Liu}(2020)}]{Fang_No-Go}%
  \BibitemOpen
  \bibfield  {author} {\bibinfo {author} {\bibfnamefont {K.}~\bibnamefont
  {Fang}}\ and\ \bibinfo {author} {\bibfnamefont {Z.-W.}\ \bibnamefont {Liu}},\
  }\href {\doibase 10.1103/PhysRevLett.125.060405} {\bibfield  {journal}
  {\bibinfo  {journal} {Phys. Rev. Lett.}\ }\textbf {\bibinfo {volume} {125}},\
  \bibinfo {pages} {060405} (\bibinfo {year} {2020})}\BibitemShut {NoStop}%
\bibitem [{\citenamefont {Modi}\ \emph {et~al.}(2012)\citenamefont {Modi},
  \citenamefont {Brodutch}, \citenamefont {Cable}, \citenamefont {Paterek},\
  and\ \citenamefont {Vedral}}]{Modi_2012}%
  \BibitemOpen
  \bibfield  {author} {\bibinfo {author} {\bibfnamefont {K.}~\bibnamefont
  {Modi}}, \bibinfo {author} {\bibfnamefont {A.}~\bibnamefont {Brodutch}},
  \bibinfo {author} {\bibfnamefont {H.}~\bibnamefont {Cable}}, \bibinfo
  {author} {\bibfnamefont {T.}~\bibnamefont {Paterek}}, \ and\ \bibinfo
  {author} {\bibfnamefont {V.}~\bibnamefont {Vedral}},\ }\href {\doibase
  10.1103/RevModPhys.84.1655} {\bibfield  {journal} {\bibinfo  {journal} {Rev.
  Mod. Phys.}\ }\textbf {\bibinfo {volume} {84}},\ \bibinfo {pages} {1655}
  (\bibinfo {year} {2012})}\BibitemShut {NoStop}%
\bibitem [{\citenamefont {Streltsov}\ \emph {et~al.}(2017)\citenamefont
  {Streltsov}, \citenamefont {Adesso},\ and\ \citenamefont
  {Plenio}}]{Streltsov-CoherenceRT}%
  \BibitemOpen
  \bibfield  {author} {\bibinfo {author} {\bibfnamefont {A.}~\bibnamefont
  {Streltsov}}, \bibinfo {author} {\bibfnamefont {G.}~\bibnamefont {Adesso}}, \
  and\ \bibinfo {author} {\bibfnamefont {M.~B.}\ \bibnamefont {Plenio}},\
  }\href {\doibase 10.1103/RevModPhys.89.041003} {\bibfield  {journal}
  {\bibinfo  {journal} {Rev. Mod. Phys.}\ }\textbf {\bibinfo {volume} {89}},\
  \bibinfo {pages} {041003} (\bibinfo {year} {2017})}\BibitemShut {NoStop}%
\bibitem [{\citenamefont {Gour}\ \emph {et~al.}(2015)\citenamefont {Gour},
  \citenamefont {M{\"{u}}ller}, \citenamefont {Narasimhachar}, \citenamefont
  {Spekkens},\ and\ \citenamefont {{Yunger Halpern}}}]{GOUR20151}%
  \BibitemOpen
  \bibfield  {author} {\bibinfo {author} {\bibfnamefont {G.}~\bibnamefont
  {Gour}}, \bibinfo {author} {\bibfnamefont {M.~P.}\ \bibnamefont
  {M{\"{u}}ller}}, \bibinfo {author} {\bibfnamefont {V.}~\bibnamefont
  {Narasimhachar}}, \bibinfo {author} {\bibfnamefont {R.~W.}\ \bibnamefont
  {Spekkens}}, \ and\ \bibinfo {author} {\bibfnamefont {N.}~\bibnamefont
  {{Yunger Halpern}}},\ }\href {\doibase
  https://doi.org/10.1016/j.physrep.2015.04.003} {\bibfield  {journal}
  {\bibinfo  {journal} {Phys. Rep.}\ }\textbf {\bibinfo {volume} {583}},\
  \bibinfo {pages} {1 } (\bibinfo {year} {2015})}\BibitemShut {NoStop}%
\bibitem [{\citenamefont {Goold}\ \emph {et~al.}(2016)\citenamefont {Goold},
  \citenamefont {Huber}, \citenamefont {Riera}, \citenamefont {del Rio},\ and\
  \citenamefont {Skrzypczyk}}]{Goold_2016}%
  \BibitemOpen
  \bibfield  {author} {\bibinfo {author} {\bibfnamefont {J.}~\bibnamefont
  {Goold}}, \bibinfo {author} {\bibfnamefont {M.}~\bibnamefont {Huber}},
  \bibinfo {author} {\bibfnamefont {A.}~\bibnamefont {Riera}}, \bibinfo
  {author} {\bibfnamefont {L.}~\bibnamefont {del Rio}}, \ and\ \bibinfo
  {author} {\bibfnamefont {P.}~\bibnamefont {Skrzypczyk}},\ }\href {\doibase
  10.1088/1751-8113/49/14/143001} {\bibfield  {journal} {\bibinfo  {journal}
  {J. Phys. A: Math. Theor.}\ }\textbf {\bibinfo {volume} {49}},\ \bibinfo
  {pages} {143001} (\bibinfo {year} {2016})}\BibitemShut {NoStop}%
\bibitem [{\citenamefont {Adesso}\ \emph {et~al.}(2016)\citenamefont {Adesso},
  \citenamefont {Bromley},\ and\ \citenamefont {Cianciaruso}}]{Adesso_2016}%
  \BibitemOpen
  \bibfield  {author} {\bibinfo {author} {\bibfnamefont {G.}~\bibnamefont
  {Adesso}}, \bibinfo {author} {\bibfnamefont {T.~R.}\ \bibnamefont {Bromley}},
  \ and\ \bibinfo {author} {\bibfnamefont {M.}~\bibnamefont {Cianciaruso}},\
  }\href {\doibase 10.1088/1751-8113/49/47/473001} {\bibfield  {journal}
  {\bibinfo  {journal} {J. Phys. A: Math. Theor.}\ }\textbf {\bibinfo {volume}
  {49}},\ \bibinfo {pages} {473001} (\bibinfo {year} {2016})}\BibitemShut
  {NoStop}%
\bibitem [{\citenamefont {Torun}\ and\ \citenamefont
  {Yildiz}(2018)}]{Torun-DetCoherence}%
  \BibitemOpen
  \bibfield  {author} {\bibinfo {author} {\bibfnamefont {G.}~\bibnamefont
  {Torun}}\ and\ \bibinfo {author} {\bibfnamefont {A.}~\bibnamefont {Yildiz}},\
  }\href {\doibase 10.1103/PhysRevA.97.052331} {\bibfield  {journal} {\bibinfo
  {journal} {Phys. Rev. A}\ }\textbf {\bibinfo {volume} {97}},\ \bibinfo
  {pages} {052331} (\bibinfo {year} {2018})}\BibitemShut {NoStop}%
\bibitem [{\citenamefont {Killoran}\ \emph {et~al.}(2016)\citenamefont
  {Killoran}, \citenamefont {Steinhoff},\ and\ \citenamefont
  {Plenio}}]{Plenio-Nonclassicality}%
  \BibitemOpen
  \bibfield  {author} {\bibinfo {author} {\bibfnamefont {N.}~\bibnamefont
  {Killoran}}, \bibinfo {author} {\bibfnamefont {F.~E.~S.}\ \bibnamefont
  {Steinhoff}}, \ and\ \bibinfo {author} {\bibfnamefont {M.~B.}\ \bibnamefont
  {Plenio}},\ }\href {\doibase 10.1103/PhysRevLett.116.080402} {\bibfield
  {journal} {\bibinfo  {journal} {Phys. Rev. Lett.}\ }\textbf {\bibinfo
  {volume} {116}},\ \bibinfo {pages} {080402} (\bibinfo {year}
  {2016})}\BibitemShut {NoStop}%
\bibitem [{\citenamefont {Tan}\ \emph {et~al.}(2017)\citenamefont {Tan},
  \citenamefont {Volkoff}, \citenamefont {Kwon},\ and\ \citenamefont
  {Jeong}}]{KCTan2017}%
  \BibitemOpen
  \bibfield  {author} {\bibinfo {author} {\bibfnamefont {K.~C.}\ \bibnamefont
  {Tan}}, \bibinfo {author} {\bibfnamefont {T.}~\bibnamefont {Volkoff}},
  \bibinfo {author} {\bibfnamefont {H.}~\bibnamefont {Kwon}}, \ and\ \bibinfo
  {author} {\bibfnamefont {H.}~\bibnamefont {Jeong}},\ }\href {\doibase
  10.1103/PhysRevLett.119.190405} {\bibfield  {journal} {\bibinfo  {journal}
  {Phys. Rev. Lett.}\ }\textbf {\bibinfo {volume} {119}},\ \bibinfo {pages}
  {190405} (\bibinfo {year} {2017})}\BibitemShut {NoStop}%
\bibitem [{\citenamefont {Regula}\ \emph {et~al.}(2018)\citenamefont {Regula},
  \citenamefont {Piani}, \citenamefont {Cianciaruso}, \citenamefont {Bromley},
  \citenamefont {Streltsov},\ and\ \citenamefont {Adesso}}]{Regula_2018}%
  \BibitemOpen
  \bibfield  {author} {\bibinfo {author} {\bibfnamefont {B.}~\bibnamefont
  {Regula}}, \bibinfo {author} {\bibfnamefont {M.}~\bibnamefont {Piani}},
  \bibinfo {author} {\bibfnamefont {M.}~\bibnamefont {Cianciaruso}}, \bibinfo
  {author} {\bibfnamefont {T.~R.}\ \bibnamefont {Bromley}}, \bibinfo {author}
  {\bibfnamefont {A.}~\bibnamefont {Streltsov}}, \ and\ \bibinfo {author}
  {\bibfnamefont {G.}~\bibnamefont {Adesso}},\ }\href {\doibase
  10.1088/1367-2630/aaae9d} {\bibfield  {journal} {\bibinfo  {journal} {N. J.
  Phys.}\ }\textbf {\bibinfo {volume} {20}},\ \bibinfo {pages} {033012}
  (\bibinfo {year} {2018})}\BibitemShut {NoStop}%
\bibitem [{\citenamefont {Horn}\ and\ \citenamefont
  {Johson}(2013)}]{Horn-GramMatrix}%
  \BibitemOpen
  \bibfield  {author} {\bibinfo {author} {\bibfnamefont {R.~A.}\ \bibnamefont
  {Horn}}\ and\ \bibinfo {author} {\bibfnamefont {C.~R.}\ \bibnamefont
  {Johson}},\ }\href@noop {} {\emph {\bibinfo {title} {Matrix Analysis}}},\
  \bibinfo {edition} {2nd}\ ed.\ (\bibinfo  {publisher} {Cambridge University
  Press},\ \bibinfo {year} {2013})\BibitemShut {NoStop}%
\bibitem [{\citenamefont {Genoni}\ and\ \citenamefont
  {Tufarelli}(2019)}]{Genoni_2019}%
  \BibitemOpen
  \bibfield  {author} {\bibinfo {author} {\bibfnamefont {M.~G.}\ \bibnamefont
  {Genoni}}\ and\ \bibinfo {author} {\bibfnamefont {T.}~\bibnamefont
  {Tufarelli}},\ }\href {\doibase 10.1088/1751-8121/ab3fe0} {\bibfield
  {journal} {\bibinfo  {journal} {J. Phys. A: Math. Theor.}\ }\textbf {\bibinfo
  {volume} {52}},\ \bibinfo {pages} {434002} (\bibinfo {year}
  {2019})}\BibitemShut {NoStop}%
\bibitem [{\citenamefont {Winter}\ and\ \citenamefont
  {Yang}(2016)}]{Winter120404}%
  \BibitemOpen
  \bibfield  {author} {\bibinfo {author} {\bibfnamefont {A.}~\bibnamefont
  {Winter}}\ and\ \bibinfo {author} {\bibfnamefont {D.}~\bibnamefont {Yang}},\
  }\href {\doibase 10.1103/PhysRevLett.116.120404} {\bibfield  {journal}
  {\bibinfo  {journal} {Phys. Rev. Lett.}\ }\textbf {\bibinfo {volume} {116}},\
  \bibinfo {pages} {120404} (\bibinfo {year} {2016})}\BibitemShut {NoStop}%
\bibitem [{\citenamefont {Renes}(2016)}]{Renes_RT}%
  \BibitemOpen
  \bibfield  {author} {\bibinfo {author} {\bibfnamefont {J.~M.}\ \bibnamefont
  {Renes}},\ }\href {\doibase 10.1063/1.4972295} {\bibfield  {journal}
  {\bibinfo  {journal} {J. Math. Phys.}\ }\textbf {\bibinfo {volume} {57}},\
  \bibinfo {pages} {122202} (\bibinfo {year} {2016})}\BibitemShut {NoStop}%
\bibitem [{\citenamefont {\AA{}berg}(2014)}]{Aberg_CC}%
  \BibitemOpen
  \bibfield  {author} {\bibinfo {author} {\bibfnamefont {J.}~\bibnamefont
  {\AA{}berg}},\ }\href {\doibase 10.1103/PhysRevLett.113.150402} {\bibfield
  {journal} {\bibinfo  {journal} {Phys. Rev. Lett.}\ }\textbf {\bibinfo
  {volume} {113}},\ \bibinfo {pages} {150402} (\bibinfo {year}
  {2014})}\BibitemShut {NoStop}%
\bibitem [{\citenamefont {Bu}\ \emph {et~al.}(2016)\citenamefont {Bu},
  \citenamefont {Singh},\ and\ \citenamefont {Wu}}]{Kaifeng_CC}%
  \BibitemOpen
  \bibfield  {author} {\bibinfo {author} {\bibfnamefont {K.}~\bibnamefont
  {Bu}}, \bibinfo {author} {\bibfnamefont {U.}~\bibnamefont {Singh}}, \ and\
  \bibinfo {author} {\bibfnamefont {J.}~\bibnamefont {Wu}},\ }\href {\doibase
  10.1103/PhysRevA.93.042326} {\bibfield  {journal} {\bibinfo  {journal} {Phys.
  Rev. A}\ }\textbf {\bibinfo {volume} {93}},\ \bibinfo {pages} {042326}
  (\bibinfo {year} {2016})}\BibitemShut {NoStop}%
\bibitem [{\citenamefont {Bhatia}(1997)}]{Bhatia}%
  \BibitemOpen
  \bibfield  {author} {\bibinfo {author} {\bibfnamefont {R.}~\bibnamefont
  {Bhatia}},\ }\href@noop {} {\emph {\bibinfo {title} {Matrix Analysis}}}\
  (\bibinfo  {publisher} {Springer-Verlag},\ \bibinfo {year}
  {1997})\BibitemShut {NoStop}%
\bibitem [{\citenamefont {Marshall}\ \emph {et~al.}(2009)\citenamefont
  {Marshall}, \citenamefont {Olkin},\ and\ \citenamefont {Arnold}}]{Marshall}%
  \BibitemOpen
  \bibfield  {author} {\bibinfo {author} {\bibfnamefont {A.~W.}\ \bibnamefont
  {Marshall}}, \bibinfo {author} {\bibfnamefont {I.}~\bibnamefont {Olkin}}, \
  and\ \bibinfo {author} {\bibfnamefont {B.~C.}\ \bibnamefont {Arnold}},\
  }\href@noop {} {\emph {\bibinfo {title} {Inequalities: Theory of Majorization
  and Its Applications}}},\ \bibinfo {edition} {2nd}\ ed.\ (\bibinfo
  {publisher} {Springer, New York},\ \bibinfo {year} {2009})\BibitemShut
  {NoStop}%
\end{thebibliography}
\end{document}